\documentclass[%
superscriptaddress,
nofootinbib,
 amsmath,amssymb,
 aps, pra, showpacs,
 longbibliography,
floatfix,
reprint
]{revtex4-1}
\usepackage[export]{adjustbox}

\usepackage{booktabs} % enable toprule and bottomrule
\usepackage{graphicx,xcolor}% Include figure files
\usepackage{dcolumn}% Align table columns on decimal point
\usepackage{bm}% bold math
\usepackage[breaklinks=true,colorlinks,citecolor=blue,linkcolor=blue,urlcolor=blue]{hyperref}% add hypertext capabilities
\usepackage[mathlines]{lineno}% Enable numbering of text and display math
\usepackage{mathtools}
\usepackage{tikz}

\usepackage[caption=false]{subfig}

\begin{document}

\title{Universal Effectiveness of High-Depth Circuits in Variational Eigenproblems}

\author{Joonho Kim}
\thanks{All the authors contributed equally}
\affiliation{School of Natural Sciences, 
Institute for Advanced Study, Princeton, NJ 08540, USA}
\email[Correspondence to: ]{ joonhokim@ias.edu}

\author{Jaedeok Kim}
\thanks{All the authors contributed equally}
\affiliation{
AI Center, Samsung Research, Seoul 06765, Republic of Korea}%
% \email{jd05.kim@samsung.com}

\author{Dario Rosa}
\thanks{All the authors contributed equally}
\affiliation{School of Physics, Korea Institute for Advanced Study, 
85 Hoegi-ro, Dongdaemun-gu, Seoul 02455, Republic of Korea}%
\affiliation{Department of Physics, Korea Advanced Institute of Science and Technology,\\
291 Daehak-ro, Yuseong-gu, Daejeon 34141, Republic of Korea}%

\begin{abstract}
We explore the effectiveness of variational quantum circuits in simulating the ground states of quantum many-body Hamiltonians. We show that generic high-depth circuits, performing a sequence of layer unitaries of the same form, can accurately approximate the desired states. We demonstrate their universal success by using two Hamiltonian systems with very different properties: the transverse field Ising model and the Sachdev-Ye-Kitaev model. The energy landscape of the high-depth circuits has a proper structure for the gradient-based optimization, \textit{i.e.} the presence of local extrema -- near any random initial points -- reaching the ground level energy. We further test the circuit's capability of replicating random quantum states by minimizing the Euclidean distance.

% \color{orange}
% We explore the effectiveness of high-depth, noiseless, parameteric quantum circuits by challenging their capability to simulate the ground states of quantum many-body Hamiltonians. Even a generic layered circuit Ansatz can approximate the ground state with high precision, as long as the circuit depth exceeds a certain threshold level that exponentially scales with the number of qubits, despite the abundance of the barren plateaus. This success is due to the fact that the energy landscape in the high-depth regime has a suitable structure for the gradient-based optimization, \textit{i.e.} the presence of local extrema -- near any random initial points -- reaching the ground level energy. We check if these advantages are preserved across different Hamiltonians, by working out two variational eigensolver problems for the transverse field Ising model as well as for the Sachdev-Ye-Kitaev model. 
% We expect that the contributing factors to the universal success of the high-depth circuits may also serve as the evaluation guidelines for more realistic circuit designs under hybrid quantum-classical algorithms.
\end{abstract}

\maketitle

%\tableofcontents
\setlength{\parskip}{0.3em}
\renewcommand{\thesubsection}{\arabic{subsection}}

\section{Introduction}
\label{sec:intro}
Variational Quantum Eigensolver (VQE) \cite{2014VQE,2016VQE} is one of the most promising hybrid quantum-classical (HQC) algorithms, which may offer a precise approximation of the ground state of quantum systems. It is based on the iterative application of the following three steps: state preparation, measurement and optimization. Let us briefly describe each step. First, the preparation of a trial state $|\psi(\bm{\theta})\rangle$ is carried out by successive application of unitary quantum gates that depend on variational parameters $\bm{\theta}$. Second, the measurement step estimates the trial state mean energy, 
\begin{align}
\label{eq:E_mean_def}
    E(\bm{\theta}) \equiv \langle\psi(\bm{\theta})|\mathcal{H}|\psi(\bm{\theta})\rangle,
\end{align} 
by taking the expectation value of Hamiltonian $\mathcal{H}$ of the target system over the trial state. Third, the optimization step adjusts the variational parameters $\bm{\theta}$ of the trial wavefunction to minimize the mean state energy $E(\bm{\theta})$ by applying a classical optimization algorithm. See \cite{2019VQEReview} and references therein for more details.
After a sufficient number of iterations, the variational state $|\psi(\bm{\theta}^*)\rangle$ at a convergence point $\bm{\theta} = \bm{\theta}^*$ is expected to reproduce well the ground state of the target Hamiltonian $\mathcal{H}$, under the assumption that the state Ansatz $|\psi(\bm{\theta})\rangle$ is \emph{parametrically expressible} and \emph{well-trainable} under the gradient based optimization. It is then a crucial question to find such an Ansatz. 

Ideally, we are looking for a universally effective Ansatz capable of solving the VQE problem associated with an \emph{arbitrary} target Hamiltonian. Note that it is generically a very challenging task: unless the target Hamiltonian is local, the ground states of non-local interacting systems are relatively close to typical quantum states that constitute most of the Hilbert space. Universality is thus equivalent to demand a circuit Ansatz to approximate \emph{any} random state $|\phi\rangle$ at certain values of the circuit parameters.

The existence of a parameter set ${\bm{\varphi}}$, such that $\vert \psi({\bm{\varphi}}) \rangle \simeq \vert \phi\rangle $, is therefore a necessary condition for the effectiveness of the variational circuit. On top of that, we must concern if the gradient descent optimization can actually reach the parameters ${\bm{\varphi}}$ from randomly initialized ones. Suppose that we use a layered quantum circuit composed of repetitive application of variational layers with the same architecture. Since increasing the number of layers extends the dimension of the parameter space, it can affect the circuit's expressibility only positively. However, there has been a reported tension between increasing depth of layers and trainability of the circuit via minimizing the mean energy, \eqref{eq:E_mean_def}, known as the barren plateau phenomenon \cite{McClean2018BarrenPI}. 

%however not sufficient but merely necessary for universal effectiveness of the circuit. 
% One also needs to look at whether or not the gradient descent optimization can actually reach the parameters ${\bm{\varphi}}$.

When the layered circuit reaches a certain depth such that it evolves to an approximate 2-design, a numerical experiment \cite{McClean2018BarrenPI} has shown the exponential decay of the variance of energy gradients $\nabla_{\bm{\theta}} E(\bm{\theta})$ with respect to the number $n$ of qubits --- for random circuit states obtained by uniformly sampling from the parameter space. Combined with another fact that the random energy gradient vanishes on average, Chebyshev’s inequality implies that the initial energy gradient can exponentially rarely deviate from zero. Such diminishing gradients hinder the beginning of efficient energy minimization, possibly causing the variational state to be stuck on non-optimal plateaus.

There have been several proposals to overcome the vanishing gradient problem and optimize the variational circuit. The most obvious approach is to incorporate some physics information on the target Hamiltonian \cite{2019ANSATZ_BARR_SPIN,2020ANSATZ_SYMM_BARR,2020SYMM_VQE_OPTIM,2020HVA}, \textit{e.g.} the ground state symmetry, for designing a less generic problem-tailored Ansatz.
% , though it is not always readily obtainable. 
As an alternative direction towards the universal applicability of the variational algorithms, novel initialization \cite{2019INIT_BARR,2019PRELEARN_BARR}, architecture \cite{2020COST_BARR,Volkoff2020LargeGV}, and optimization  \cite{2019NaturalGrad, 2020LAYER_BARR, Koczor2020} of generic purpose circuits have been developed to enhance the circuit performance.

The main goal of this paper is to demonstrate the inherent capability of variational circuits as universal and accurate eigensolvers for generic Hamiltonians -- if they can even approximate close-to-random states. To this end, we will simply put aside the barren plateau problem by sufficiently increasing the classical computation capacity. Specifically, we will consider the high-depth regime of the layered quantum circuit, thus building an exponentially high-dimensional parameter space. In this regime, the variational circuit has sufficiently many layers and consequently many parameters. Hence, the norm of the gradient vector $\Vert \nabla_{\bm{\theta}} E(\bm{\theta}) \Vert$ can still grow to a finite size, 
capable of moving the circuit parameter ${\bm \theta}$ from initial points,
even though the magnitudes of the individual components $|\partial_i E(\bm{\theta})|$ are exponentially suppressed for the number $n$ of qubits.
We will illustrate the effectiveness of the high-depth circuits by solving two concrete VQE problems for the following quantum many-body Hamiltonians: the 1d Ising model in a constant transverse magnetic field, which is a prototypical model of locally interacting spin-chain systems, and the Sachdev-Ye-Kitaev (SYK) model, which is a strongly interacting quantum mechanical system of Majorana fermions \cite{1993SY,2015Kitaev,Maldacena:2016hyu}. Despite the striking contrast in these two Hamiltonians' ground state properties, we will find that the high-depth circuit achieves a very high fidelity in approximating both ground states. Moreover, we will see that the high-depth circuits can narrow the Euclidean distance from random states up to arbitrary precision, indicating outstanding expressibility and trainability that stems from the high-dimensionality of  ${\bm \theta}$.

The remarkable efficiency of the gradient descent optimization in the \emph{over-parameterized} regime was also reported for approximating random unitary matrices \cite{2020UNITARY} and ground states \cite{2020HVA} with certain variational Ansatzes. More generally, over-parameterization is an active research topic that underlies the success of deep learning. It makes large neural networks capable to reach a global minimum during the optimization process, despite the non-convexity of the energy landscape \cite{belkin2019reconciling,liu2020theory}. While searching the ground state via the energy minimization, we will observe some phenomena similar to what happens during the training of the neural networks with large parameter spaces, summarized as follows:

First, the energy landscape looks fairly simple in the local vicinity of randomly chosen points \cite{Koczor2020}. Generically, an initial point is already confined in a certain basin of attraction, such that an emanating optimization trajectory can quickly arrive at a nearby local extremum. Especially if the circuit has enough layers to become an approximate 2-design, almost all uniformly sampled initial states end up with rather homogeneous energy levels. 
% The finally converged states after the VQE calculus are also less variable, resulting in a similar energy level and local curvature in the parameter space.

Second, for high-depth circuits, all the local extrema in the energy landscape, reachable by the VQE optimization from randomly initialized points, are substantially close to the exact ground energy, i.e., the value of the global minimum. These minima are not isolated individual points but develop multiple flat directions. It explains the robust success of the high-depth circuits in solving the VQE problems.

The rest of this paper is organized as follows. Section~\ref{sec:expressibility} introduces the architecture of the variational circuits used throughout this paper.  Section~\ref{sec:trainability} first reviews the occurrence of the barren plateau phenomenon, then examines the optimization of the variational circuits using the VQE example of the 1d Ising model coupled to a uniform transverse magnetic field. Section~\ref{sec:syk} studies the same VQE problem for the SYK model, thus showing the universal effectiveness of the high-depth circuit. 
Section~\ref{sec:euclidean} addresses the ability of the variational circuits to reproduce typical states by minimizing the Euclidean distance. In particular, it shows that the high-depth Ansatz is highly expressible and trainable, being able to reach \textit{any} random state.
Finally, Section~\ref{sec:discussion} concludes with discussions.

\section{Circuit Ansatzes}
\label{sec:expressibility}

\begin{figure}[t]
\subfloat[\label{subfig:a} Circuit Ansatz]{
    \includegraphics[width=0.95\linewidth]{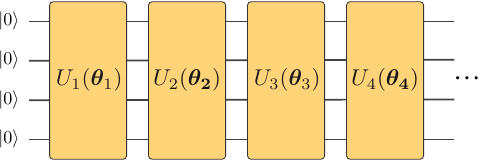}
}
\hfill
\subfloat[\label{subfig:b} Layer]{
    \includegraphics[width=0.84\linewidth]{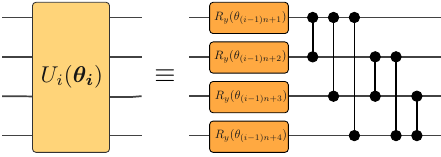}
    {\ }
}
% \hfill
% \subfloat[\label{subfig:c} even layers ]{
% \includegraphics[width=0.8\linewidth]{fig/circuit_ansatz_even.pdf}
% }
\caption{
    The layered circuit Ansatz $\vert \psi(\bm{\theta}) \rangle$ used in this paper.
}
\label{fig:circuit_ansatz}
\end{figure}

Our focus in this work is to demonstrate the efficiency of high-depth layered circuits in a typical VQE problem,
\textit{i.e.} to approximate the ground state of a given Hamiltonian. To this purpose, here we specify the architecture of the variational circuit used in our numerical experiments. Our circuit state $|\psi(\bm{\theta})\rangle$ is composed of $L$ unitary layers acting sequentially on the initial product state $|0\rangle$, \textit{i.e.}, 
\begin{align}
    \label{eq:circuit-ansatz}
    |\psi(\bm{\theta})\rangle = U_{L} (\bm{\theta}_L)\,
    U_{L-1} (\bm{\theta}_{L-1}) \cdots 
    U_{2} (\bm{\theta}_2)\, U_{1} (\bm{\theta}_1) |0\rangle,
\end{align}
where each layer $U_i(\bm{\theta}_i)$ has single-qubit $y$-rotation gates, parameterized by $n$ periodic variables $\bm{\theta}_i$,
% $\theta_{i, 1}, \cdots, \theta_{i, n}$, 
and controlled-$z$ gates operating on all pairs of $n$ qubits. More precisely, once the single-qubit $RY$ gates have acted upon all individual qubits, the entangling $CZ$ gates acting on $a$-th controlling and $b$-th targeting qubits are arranged for every integer pair $(a,b)$ satisfying $1 \leq a<b \leq n$. We have drawn the circuit state \eqref{eq:circuit-ansatz} with $n=4$ qubits in Figure~\ref{fig:circuit_ansatz}.

Note that we have deliberately chosen an unbiased circuit architecture, instead of embedding known properties of the ground states that are under search. Nonetheless, as we will see in Sections~\ref{sec:trainability},~\ref{sec:syk}~and~\ref{sec:euclidean}, the above circuit can solve the ground states of distinct Hamiltonians by minimizing $E(\bm{\theta})$ and accurately approximate random quantum states by minimizing the Euclidean distance, as long as the number $L$ of layers is sufficiently large.

% as %to be all-to-all 
% in the following %$p$-shifted 
% lexicographic order,
% \begin{align}
%     (a, b) \prec  (c, d) \quad &\text{if } a + p < c + p \ \text{ or}\\&
%     \text{if } a + p = c + p  \text{ and } b + p < d + p ,\nonumber
% \end{align}
% where $a,b,c,d,p \in \mathbb{Z} \text{ (mod $n$)} $ whose residue classes are represented by non-negative integers  $\{1,2,\cdots,n\}$. Here the inequality $(a, b) \prec  (c, d)$ means the $CZ$ gate on $(a, b)$ qubits operates prior to the $CZ$ gate on $(c, d)$ qubits. The shifting integer $p\in \mathbb{Z}_n $ of the above lexicography is 
% \begin{align}
%     p = \begin{dcases}
%     n & \text{for odd layers} \\
%     n/2 & \text{for even layers}.
%     \end{dcases}
% \end{align}
% In general, the rule to create the set of $CZ$ gates goes as follows:
% \begin{itemize}
%     \item For odd layers, we start from the first qubit and we connect it with all the other subsequent qubits, in order. We then move to the second qubit and we connect it, in order, with all the other subsequent qubits.
%     \item For even layers, we repeat exactly the same procedure but shifted, periodically, by $n / 2$.
% \end{itemize}

\section{Looking into VQE Trajectories}
\label{sec:trainability}

This section is devoted to the detailed investigation of the VQE optimization procedure, with a particular focus on the effectiveness of the high-depth circuit Ansatz \eqref{eq:circuit-ansatz}.
Our exploration will be based on a concrete Hamiltonian system, commonly used in measuring the performance of variational circuits, \textit{i.e.} the 1d Ising model in a transverse and uniform magnetic field \cite{2020Expressibility,2020HVA}.

The 1d transverse field Ising model is defined over a spin lattice of length $n$, consisting of the spin-spin coupling between nearest neighbors in the $z$ direction, as well as the spin interaction with a background uniform magnetic field along the transverse $x$ direction. Assuming periodic boundary conditions, $\sigma_{n+1}^z \equiv \sigma_{1}^z$, the ferromagnetic Ising Hamiltonian reads 
\begin{align}
    \label{eq:ham_ising}
    \mathcal{H} \equiv - \sum_{i=1}^n \sigma_i^z \sigma_{i + 1}^z - g \sum_{i} \sigma_i^x \ ,
\end{align}
where $\sigma_i^{x, y, z}$ is the Pauli operator acting on the $i$'th spin, and $g$ denotes the strength of the uniform magnetic field. 

The physics of this model has been well-studied \cite{sachdev_2011}. When the lattice size scales up to infinity, $n \to \infty$, the system undergoes a \textit{quantum} phase transition at $|g| = 1$  between the ordered ($|g| < 1$) and disordered ($|g| > 1$) phases. 
The former phase has the spin-flip $\mathbb{Z}_2$ symmetry that connects the two opposite ferromagnetic ground states, while the latter phase has a unique ground state, with all the spins aligned along the $x$ direction. 

We will apply the VQE algorithm to the finite $n$ Ising system, which exhibits some differences from the thermodynamic limit.  Specifically, the $\mathbb{Z}_2$ degeneracy in the $0 < |g| < 1$ phase is broken at finite $n$. Our target state will be always non-degenerate and gapped at $|g| \neq 0$. All concrete calculations presented in this section  have been done for $g=2$.

\subsection{Barren Plateaus and Classical Resolution}
\label{subsec:vanishing_grad}

% \begin{figure*}[t]
%     \centering
%     \subfloat[Sample variance of the energy derivative $\partial_k E(\bm{\theta})$]{
%         \centering
%         \label{subfig:bp-a}
%         \includegraphics[width=0.45\linewidth]{fig/ising_bp_grad_var.pdf}
%     }
%     % \newline
%     \ \ \ 
%     \subfloat[Sample average of the Euclidean norm of $\nabla_{\bm \theta} E(\bm{\theta})$]{
%         \centering
%         \label{subfig:bp-b}
%         \includegraphics[width=0.45\linewidth]{fig/ising_bp_grad_norm.pdf}
%     }
%     \caption{
%         The barren plateau experiment for the Ising Hamiltonian \eqref{eq:ham_ising}. The shades indicate (a) the variance across gradient components $\{\partial_k E(\bm{\theta})\}_{k=1}^{nL}$, (b) the first and third sample quantiles.
%     }
%     \label{fig:bp}
% \end{figure*}

\begin{figure}[t]
    \centering
        \includegraphics[width=\linewidth]{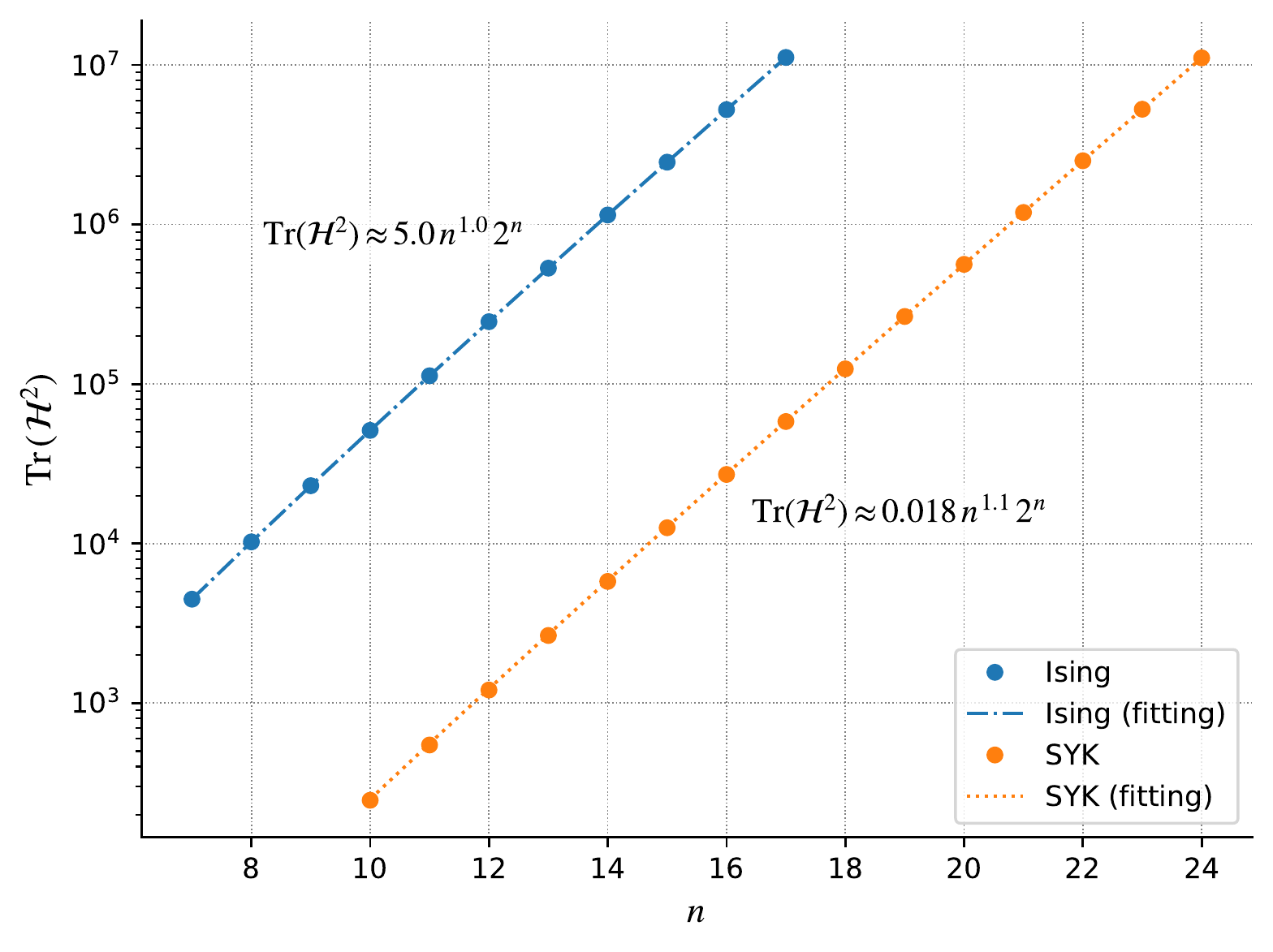}
    \caption{
        $\mathrm{Tr}(\mathcal H^2)$ of the Ising model \eqref{eq:ham_ising} and the SYK model \eqref{eq:ham_syk} as a function of the system size $n$. The dashed lines are the regression lines with the functional form of $a \cdot n^b \, 2^n$, based on the numerical data denoted as small circles.
    }
    \label{fig:spectral-density}
\end{figure}

It was argued in \cite{McClean2018BarrenPI} that optimization of the quantum variational circuits under the random parameter initialization comes with an inherent difficulty, given by the problem of vanishing gradients. 
For the circuit unitary ensemble that is quantum 2-design, random initial parameters are typically located on a plateau in the energy landscape.
It means that the gradient-based optimization cannot even roll out. The odds of having non-vanishing gradients at random points decays exponentially with the system size $n$, impeding a large-scale application of the variational circuits. Its consequence would be even more detrimental on actual quantum devices, where the sampling noise must be taken into account \cite{McClean2018BarrenPI,2020LAYER_BARR}. 
%  Accurate computation of a gradient which decays exponentially with the system size $n$ demands exponentially many samples, eliminating the scaling advantage of quantum machines . 
Given its prominence, we begin by reviewing the vanishing gradient problem in the VQE setting.

Consider the ensemble of the energy gradients over the parameter space of the variational circuit in Figure~\ref{fig:circuit_ansatz}. For analyzing the $k$'th component, $\partial_{k} E(\bm{\theta})$, that belongs to the $\ell$'th variational layer, it is convenient to group the $L$ layer unitaries \eqref{eq:circuit-ansatz} into the following two blocks:
% Let us collect the energy gradient at every point in the $nL$-dimensional torus, $[0, 2\pi)^{\otimes nL}$, for the VQE problem with the layered circuit \eqref{eq:circuit-ansatz} and target Hamiltonian \eqref{eq:ham_ising}. 
\begin{align}
    |\psi(\bm{\theta})\rangle = U_{-} (\bm{\theta}_-)\,
    %  \,
    U_{+} (\bm{\theta}_+)\, |0\rangle,
\end{align}
where $\bm{\theta}_q = \{\theta_{p}\, \vert\, \lfloor\frac{p}{n}\rfloor = q\}$, $\bm{\theta}_{+} = \bigcup_{a \leq \ell} \bm{\theta}_a$, $\bm{\theta}_{-} = \bigcup_{a > \ell} \bm{\theta}_a$, 
\begin{equation}
\begin{split}
    U_+(\bm{\theta}_+) &\equiv U_{\ell} (\bm{\theta}_{\ell}) U_{\ell-1} (\bm{\theta}_{\ell-1}) \cdots U_{2} (\bm{\theta}_{2}) \,U_{1} (\bm{\theta}_{1}), \\
    U_-(\bm{\theta}_-) &\equiv U_{L} (\bm{\theta}_{L}) \cdots U_{\ell+2} (\bm{\theta}_{\ell+2}) \,U_{\ell+1} (\bm{\theta}_{\ell+1}).
\end{split}
\end{equation}  
As the partial derivative $\partial_k$ acts only  on $U_{\ell} (\bm{\theta}_{\ell})$, the  variance of the $k$'th gradient component, $\text{Var}_{\bm{\theta}}[\partial_{k} E(\bm{\theta})]$, is 
\begin{align}
\label{eq:bp-var}
    -\int \frac{d\bm{\theta}}{(2\pi)^{nL}} \langle\psi(\bm{\theta}) | [ U_-(\bm{\theta}_-) V_k U_-(\bm{\theta}_-)^\dagger, \mathcal{H}]| \psi(\bm{\theta}) \rangle^2
\end{align}
where $V_k$ denotes the Pauli operator $\sigma^y$ acting on the $k$'th spin variable, such that $\text{Tr}(V_k) = 0$ and $\text{Tr}(V_k^2) = 2^n$. 

\begin{figure*}[t!]
    \centering
    \subfloat[Sample variance of the energy derivative $\partial_k E(\bm{\theta})$]{
        \centering
        \label{subfig:bp-a}
        \includegraphics[width=0.45\linewidth]{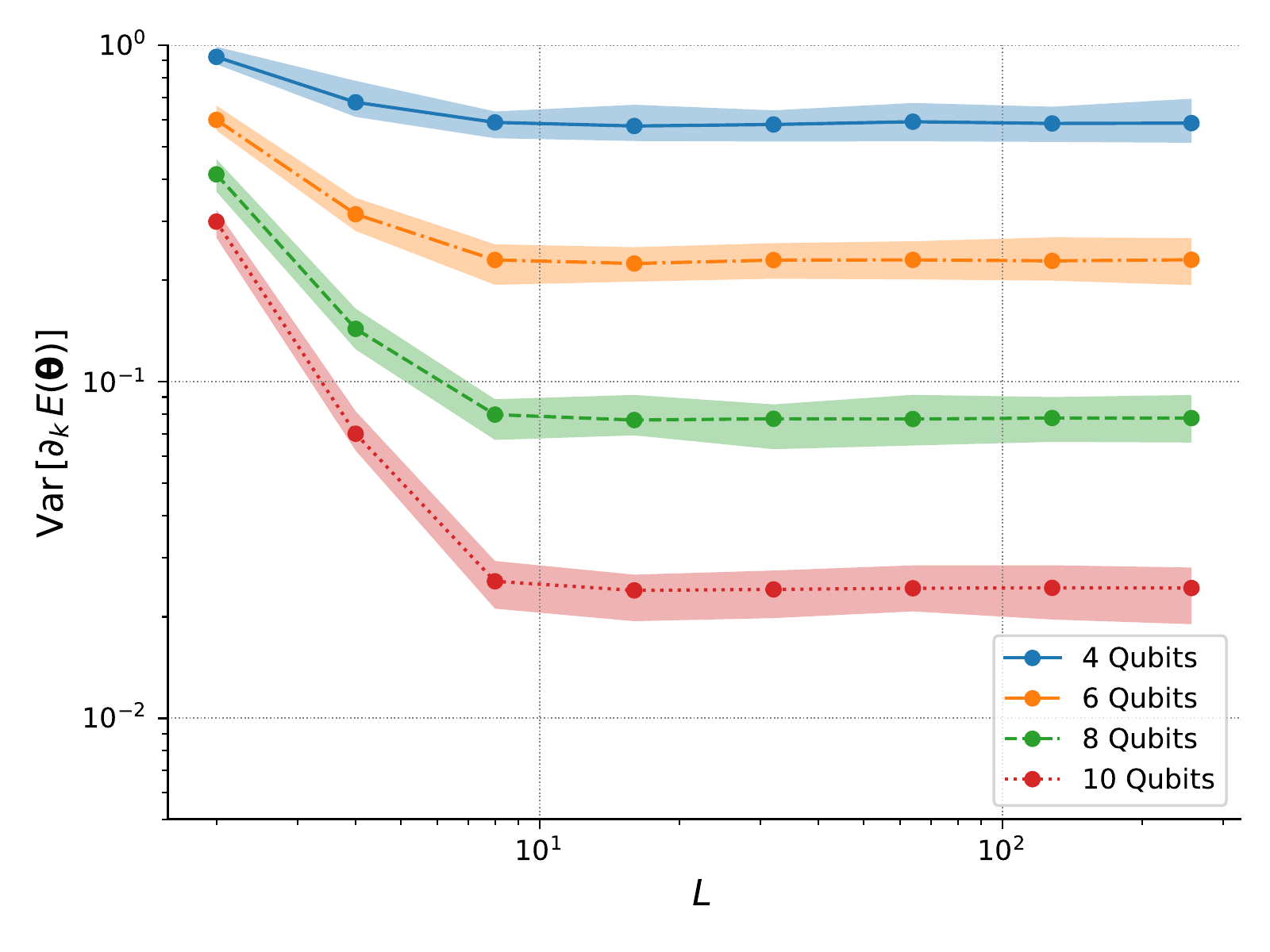}
    }
    % \newline
    \ \ \ 
    \subfloat[Sample average of the Euclidean norm of $\nabla_{\bm \theta} E(\bm{\theta})$]{
        \centering
        \label{subfig:bp-b}
        \includegraphics[width=0.45\linewidth]{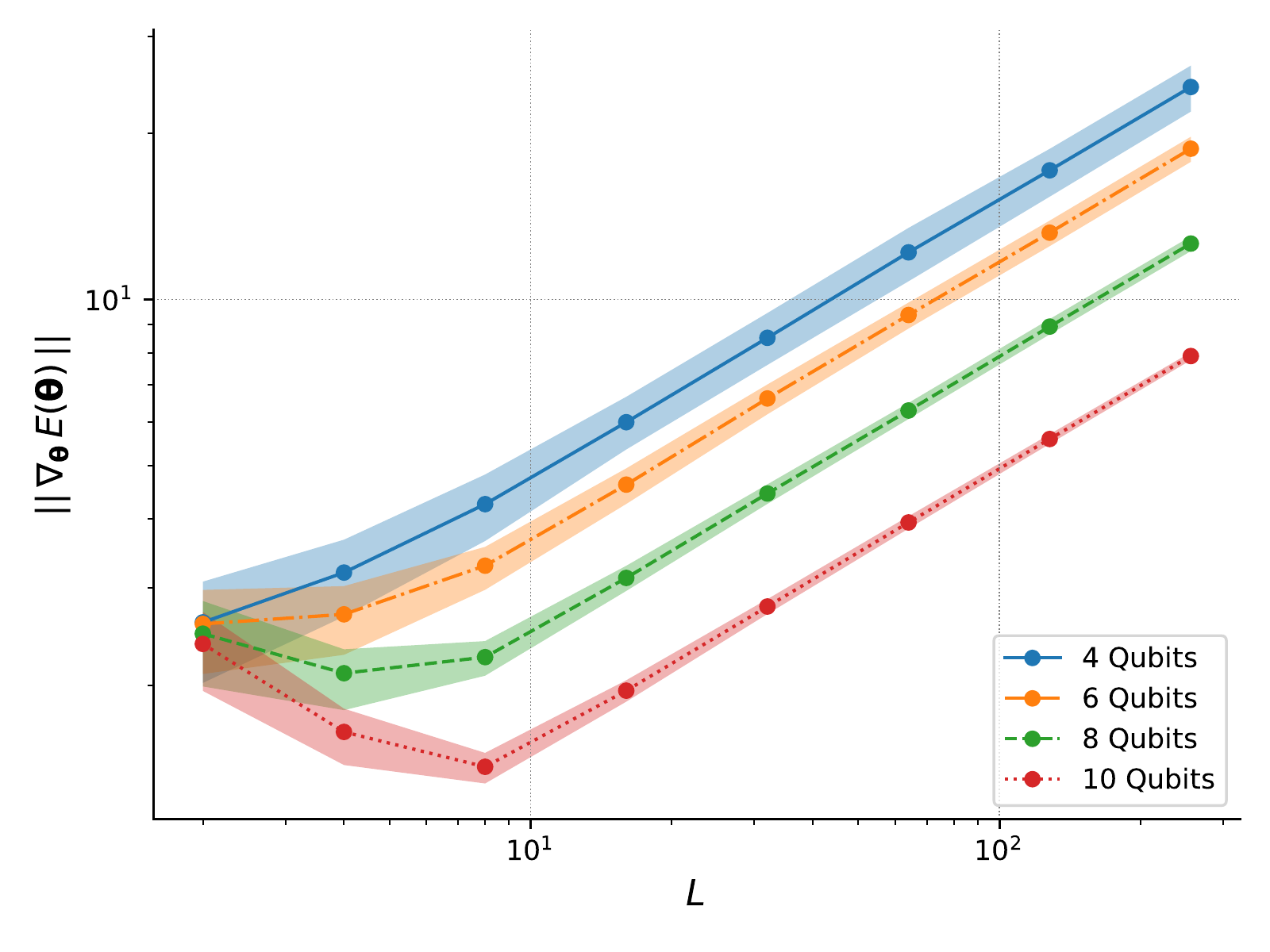}
    }
    \caption{
        The barren plateau experiment for the Ising Hamiltonian \eqref{eq:ham_ising}. The shades indicate (a) the variance across gradient components $\{\partial_k E(\bm{\theta})\}_{k=1}^{nL}$, (b) the first and third sample quantiles.
    }
    \label{fig:bp}
\end{figure*} 

If we assume the quantum 2-design property of $U_+(\bm{\theta}_+)$ and/or $U_-(\bm{\theta}_-)$, the above integral \eqref{eq:bp-var} can be replaced with the unitary matrix integral. In that case, the matrix integral can be handled exactly and simplified to 
\begin{align}
    \label{eq:bp-case1}
      \frac{2\text{Tr}(\mathcal{H}^2)}{2^{2n}}.
\end{align}
Such simplification \eqref{eq:bp-case1} happens when both $U_\pm(\bm{\theta}_\pm)$ are 2-designs. Instead, if the 2-design condition does not hold for either of $U_\pm(\bm{\theta}_\pm)$, we have the following expression:
\begin{align}
    \label{eq:bp-case2}
    - \frac{1}{2^{2n}}\int \frac{d\bm{\theta}_-}{(2\pi)^{n(L-\ell)}}  \,\text{Tr} ([V_k, U_-(\bm{\theta}_-)^\dagger \mathcal{H} U_-(\bm{\theta}_-)]^2)    
\end{align}
if only $U_+(\bm{\theta}_+)$ is a 2-design while $U_-(\bm{\theta}_-)$ is not, or
\begin{align}
    \label{eq:bp-case3}
     - \frac{\text{Tr}(\mathcal{H}^2)}{2^{2n}} \int \frac{d\bm{\theta}_+}{(2\pi)^{n\ell}}  \text{Tr}([V_k, U_+(\bm{\theta}_+)^\dagger \rho\, U_+(\bm{\theta}_+)]^2)
\end{align}
where $\rho = |0\rangle\langle 0|$, if not $U_+(\bm{\theta}_+)$ but only $U_-(\bm{\theta}_-)$ is a 2-design. Asymptotically in the system size $n$, the above expressions \eqref{eq:bp-case1}--\eqref{eq:bp-case3} are all bounded as
\begin{align}
\label{eq:BP-bound}
     0 \leq \text{Var}_{\bm{\theta}}[\partial_{k} E(\bm{\theta})]\leq  \frac{4\text{Tr}(\mathcal{H}^2)}{2^{2n}} \ ,
\end{align}
where we find the upper bound by expanding the commutator inside the integral of \eqref{eq:bp-case2}~and~\eqref{eq:bp-case3}, then applying the following trace inequality that holds for two Hermitian matrices $A$, $B$ \cite{Yang2002}:
\begin{align}
    |\text{Tr}(AB)^{2m}| \leq \text{Tr}(A^{2m}B^{2m})\quad  \text{for } m\in \mathbb{N}.
\end{align}

% \begin{figure}[t]
%     \centering
%         % \includegraphics[width=\linewidth]{trace_plot_SYK.pdf}
%         \includegraphics[width=\linewidth]{fig/trace_H2.pdf}
%     \caption{
%         $\mathrm{Tr}(\mathcal H^2)$ of the Ising model \eqref{eq:ham_ising} and the SYK model \eqref{eq:ham_syk} as a function of the system size $n$. The dashed lines are the regression lines with the functional form of $a \cdot n^b \, 2^n$, based on the numerical data denoted as small circles.
%     }
%     \label{fig:spectral-density}
% \end{figure}

To determine the scaling behavior of the upper bound \eqref{eq:BP-bound} with respect to the system size $n$, we examine how $\text{Tr}(\mathcal{H}^2)$ scales with $n$ by extrapolating the values obtained from exact diagonalization of the Hamiltonian \eqref{eq:ham_ising} up to $n \leq 17$. The result is presented in Figure~\ref{fig:spectral-density}. We see that the term $\text{Tr}(\mathcal{H}^2)$ scales as $2^n$, such that the exponential factor in the denominator of the upper bound \eqref{eq:BP-bound} cannot fully be balanced by $\text{Tr}(\mathcal{H}^2)$. Then, inserting the upper bound \eqref{eq:BP-bound} into Chebyshev’s inequality, which states 
\begin{align}
    \label{eq:chebyshev}
    \text{Pr}\left( |\partial_k E(\bm{\theta})| > \epsilon \right) < \frac{\text{Var}_{\bm{\theta}}[\partial_k E(\bm{\theta})]}{\epsilon^2}
\end{align}
with $\partial_k E(\bm{\theta})$ being a random variable of zero mean, the probability of having a non-zero and finite derivative is exponentially suppressed with growing $n$. Therefore, the large-scale VQE problems are expected to suffer from the vanishing gradients.

We have not yet justified the assumption that either of the unitaries $U_\pm(\bm{\theta}_\pm)$ is quantum 2-design. To see if the problem of exponentially vanishing gradients happens in the circuit unitaries of Figure~\ref{fig:circuit_ansatz}, we numerically estimate the variance of initial gradients by collecting $10^3$ random energy gradients of the Ising model. 
Figure~\ref{subfig:bp-a} clearly exhibits an exponential decay of the partial derivatives with the increasing number $n$ of qubits, where the shaded region displays component-wise fluctuations of the variance for all $1 \leq k \leq nL$. It shows that the energy landscape for the circuit in Figure~\ref{fig:circuit_ansatz} indeed contain the barren plateaus, which
hinders the circuit optimization towards the ground state.

On the other hand, Figure~\ref{subfig:bp-a} shows that the variance, at a fixed value of $n$, converges to a constant by increasing the number $L$ of layers. The variance is independent of $L$ beyond a transition point $L_0$, where the circuit unitaries with $L \geq L_0$ evolve to approximate 2-design. Due to the saturation, having exponentially many parameters can compensate for the exponential decay of individual components when calculating the gradient norm $\Vert\nabla_{\bm{\theta}} E(\bm{\theta})\Vert$, which appears in the evolution of the circuit energy under the gradient descent with an infinitesimal rate $\alpha$:
\begin{align}
    \frac{dE(\bm{\theta})}{d\tau} = -\alpha\Vert\nabla_{\bm{\theta}} E(\bm{\theta})\Vert^2
\end{align}
Therefore, the vanishing gradient problem inherent to the quantum Hilbert space can be resolved in a classical way, \textit{i.e.} by using the exponentially high-dimensional $\bm{\theta}$-space.

% Such gradient configuration behaves as a finite-sized vector in an exponentially large space, with each components being exponentially suppressed.
In agreement with the reasoning above, Figure~\ref{subfig:bp-b}
exhibits a small initial drop dominated by the transient decrease of  $\text{Var}_{\bm{\theta}}[\partial_{k} E(\bm{\theta})]$ for $L \leq L_0$ layers, then a steady increase driven by the linear growth in the number of  circuit parameters.
% the growth in the number $nL$ of the gradient components. 
One can approximate the asymptotic increase rate of the norm $\Vert\nabla_{\bm{\theta}} E(\bm{\theta})\Vert$ as 
\begin{align}
\label{eq:rate_ising}
    \Vert\nabla_{\bm{\theta}} E(\bm{\theta})\Vert \sim \sqrt{ nL \times  \text{Var}_{\bm{\theta}}[\partial_{k} E(\bm{\theta})] }
\end{align}
which agrees well with the numerically estimated growth rates between $\log \Vert\nabla_{\bm{\theta}} E(\bm{\theta})\Vert$ and $\log L$ in Figure~\ref{subfig:bp-b}: \begin{align*}
    \begin{tabular}
    {>{\centering}p{1.7cm}|*{4}{>{\centering}p{1.2cm}}}
        \hline
        \text{$n$ qubits} & 4 & 6 & 8 & 10 \cr
        \hline
        rate & 0.504 & 0.502 & 0.503 & 0.501 \cr
        % $\mathrm{SYK}_4$ & 0.503 & 0.500 & 0.479 & 0.442 \cr
        \hline
    \end{tabular}
\end{align*}

We remark that the barren plateau phenomenon concerns only the initial steps of the gradient descent update. In almost all physical systems of relevance, the Hamiltonian spectrum is symmetrically distributed around a certain value, which we canonically set to zero. A random state is therefore in a superposition of multiple Hamiltonian eigenstates, whose mean energy is almost certainly zero. On the other hand, the variational mean energy quickly becomes a negative value after a few steps of the VQE optimization, implying that the circuit state is no longer represented by sample statistics of random states.

Having found that the vanishing gradient problem can be trivially avoided at the cost of having exponentially many parameters, we turn to answer if the variational circuit with a sufficiently many layers can actually solve the VQE problems. We will examine the optimization error, the training curve, and the trajectory in the parameter space at different values of the circuit depth $L$.

\subsection{Optimizing the Circuit}
\label{subsec:speedup}

We now come back to original task of approximating the ground state of the Ising model, Eq.~\eqref{eq:ham_ising}.

The VQE optimization results of the circuit states \eqref{eq:circuit-ansatz}, with different numbers $L$ of layers and under the random initialization of the circuit parameters $\bm{\theta}$, can be summarized to the following two features: 
\begin{enumerate}
    \item For small enough $L$, the minimized energies $E(\bm{\theta^*})$ that the circuit states can reach are highly variable.
    % For small enough $L$, the initial circuit states $|\psi(\bm{\theta})\rangle$ end up with highly variable VQE energy $E(\bm{\theta^*})$.
    \item For larger $L$, the $E(\bm{\theta^*})$ distribution turns % sharply
    concentrated around a mean value which gets smaller.
\end{enumerate}
We illustrate them with the outcomes of the VQE algorithm for the case of a chain having $n=10$ sites and for $L \in \{8, 10, 12, 14, 16, 18, 20\}$ layers. To find the optimal parameters $\bm{\theta}^* $ that minimize the energy, we use the Adam optimization algorithm \cite{2015Adam}, which iteratively updates the variational parameters $\bm{\theta}$ by the exponential moving averages of gradients and their squares. It has a clear advantage in convergence speed, being widely used in a variety of deep learning models. The parameter update rule is decided by a choice of three hyperparameters $(\alpha, \beta_1, \beta_2)$. We have collected 35 independent VQE runs for the Ising Hamiltonian \eqref{eq:ham_ising}, whose initial parameters are randomly sampled from the uniform distribution  $\mathcal{U}(0, 2\pi)^{\otimes nL}$, after 500 parameter updates with the hyperparameters $(\alpha, \beta_1, \beta_2)= (0.05, 0.9, 0.999)$.

\begin{figure}[t]
    \centering
        \includegraphics[width=\linewidth]{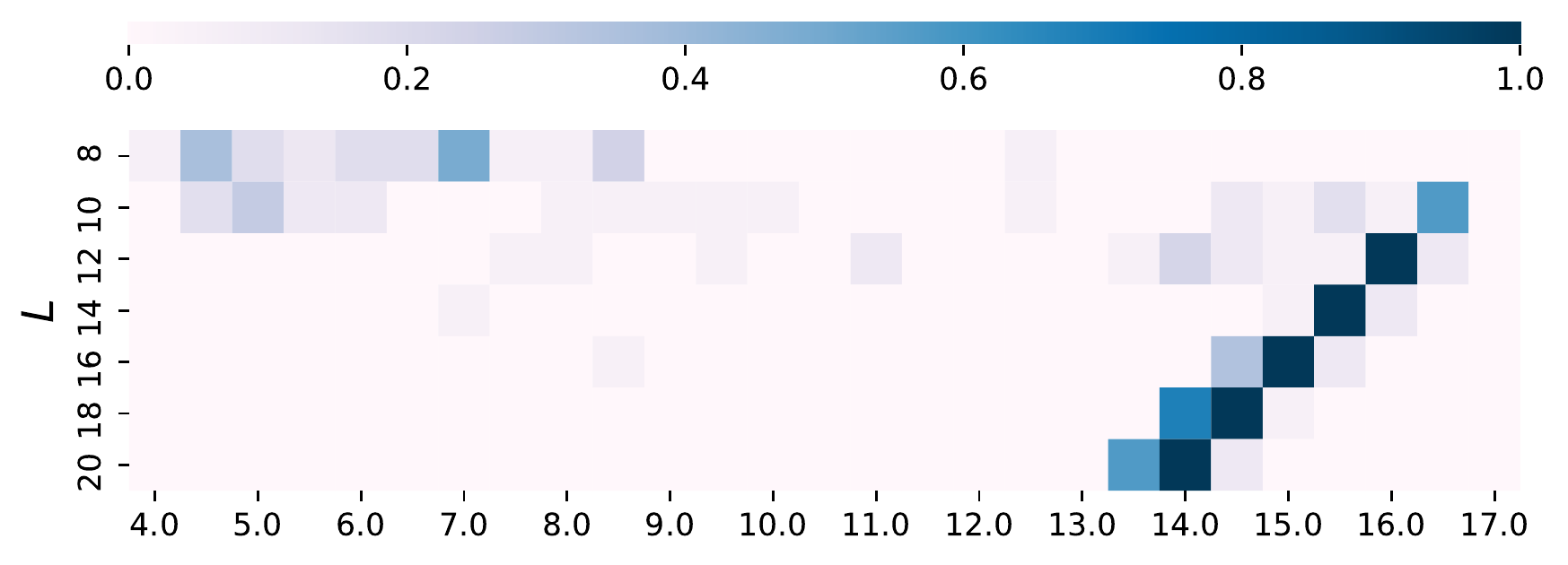}
    \caption{
        Optimized VQE energy ($E(\bm{\theta^*}) - E_0$) density for the Ising model over 35 distinct runs with random initialization.
    }
    \label{fig:vqe-ising-density}
\end{figure}
The sample distribution of the final VQE energy $E(\bm{\theta^*})$ is visualized in Figure~\ref{fig:vqe-ising-density} for different numbers $L$ of layers. One can characterize it as follows. First, the energy distribution at the local extremum $\bm{\theta^*}$ clearly exhibits the widespread  spectrum for a shallow depth, \textit{e.g.}, 
\begin{align*}
\begin{array}{cl}
    3.52 \leq E(\bm{\theta^*})  - E_{0} \leq 12.6 & \quad\text{for $L=8$},\\
    4.67 \leq E(\bm{\theta^*}) - E_0 \leq 16.9 & \quad\text{for $L=10$}.
\end{array}
\end{align*}
sometimes achieving a relatively good energy level while the gap always persists.
%, given their limited degree of the parameteric expressibility \eqref{eq:expressibility}. 
%Such variability shows that the circuit is still shallow, not having evolved yet to a random 2-design, being consistent with Figure~\ref{subfig:bp-a}.
Second, by stacking more layers, \textit{i.e.}, $L \geq 12$, 
deeper than the 2-design transition point detected in Figure~\ref{subfig:bp-a},
the energy distribution starts to concentrate around a value, which is far from the ground energy $E_0$. Third, the average value of  $E(\bm{\theta^*})$ continues to decrease for the growing number $L$ of layers, suggesting that the high-depth variational circuits can possibly simulate the ground state using the VQE optimization.

\begin{figure}[t]
    \centering
    \subfloat[The VQE optimization error $E(\bm{\theta^*}) - E_0$]{
        \centering
        \label{subfig:ising-error}
        \includegraphics[width=0.9\linewidth]{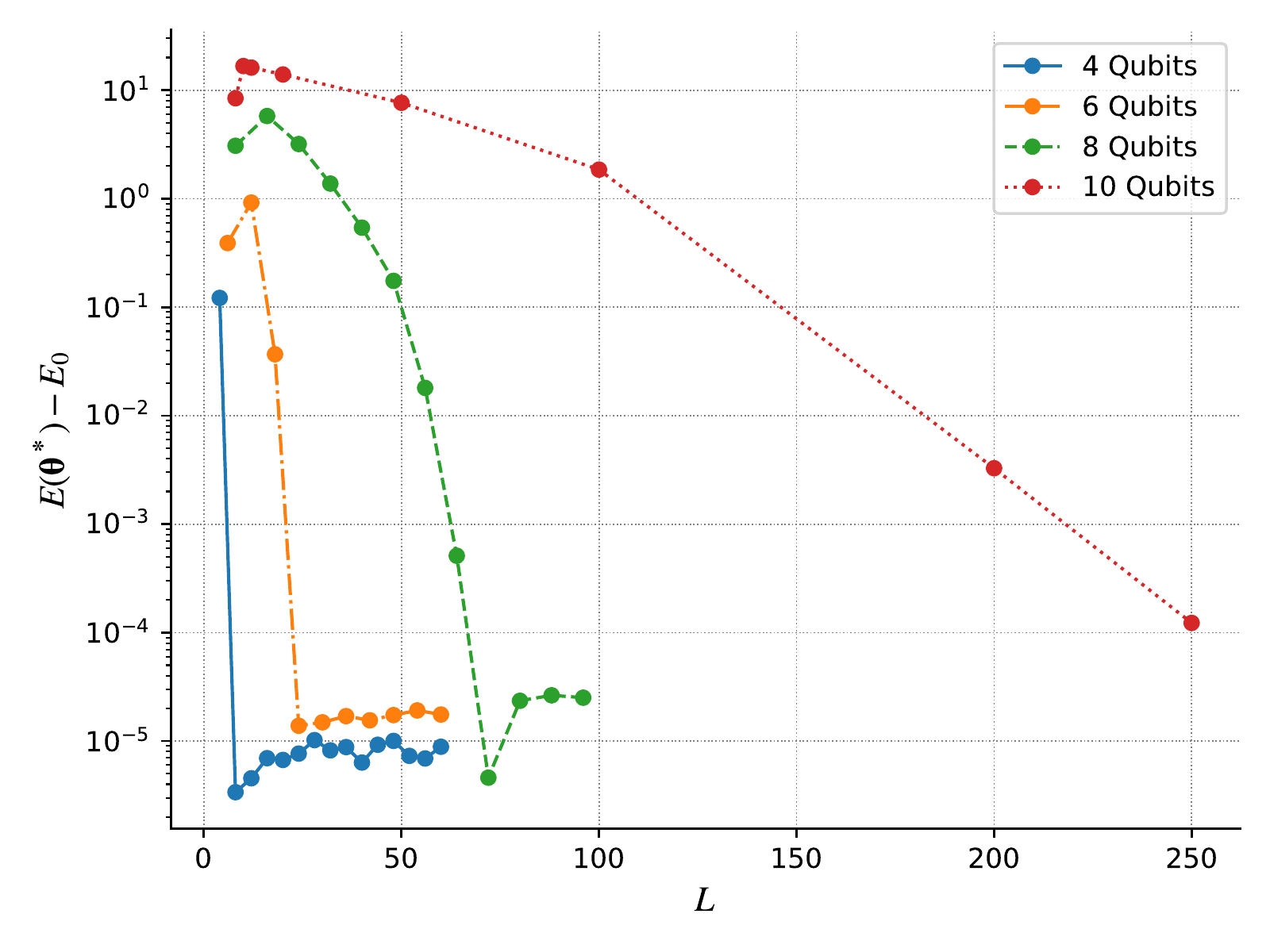}
    }
    \newline
    \subfloat[Fidelity between the optimized and true ground states]{
        \centering
        \label{subfig:ising-fidelity}
        \includegraphics[width=0.9\linewidth]{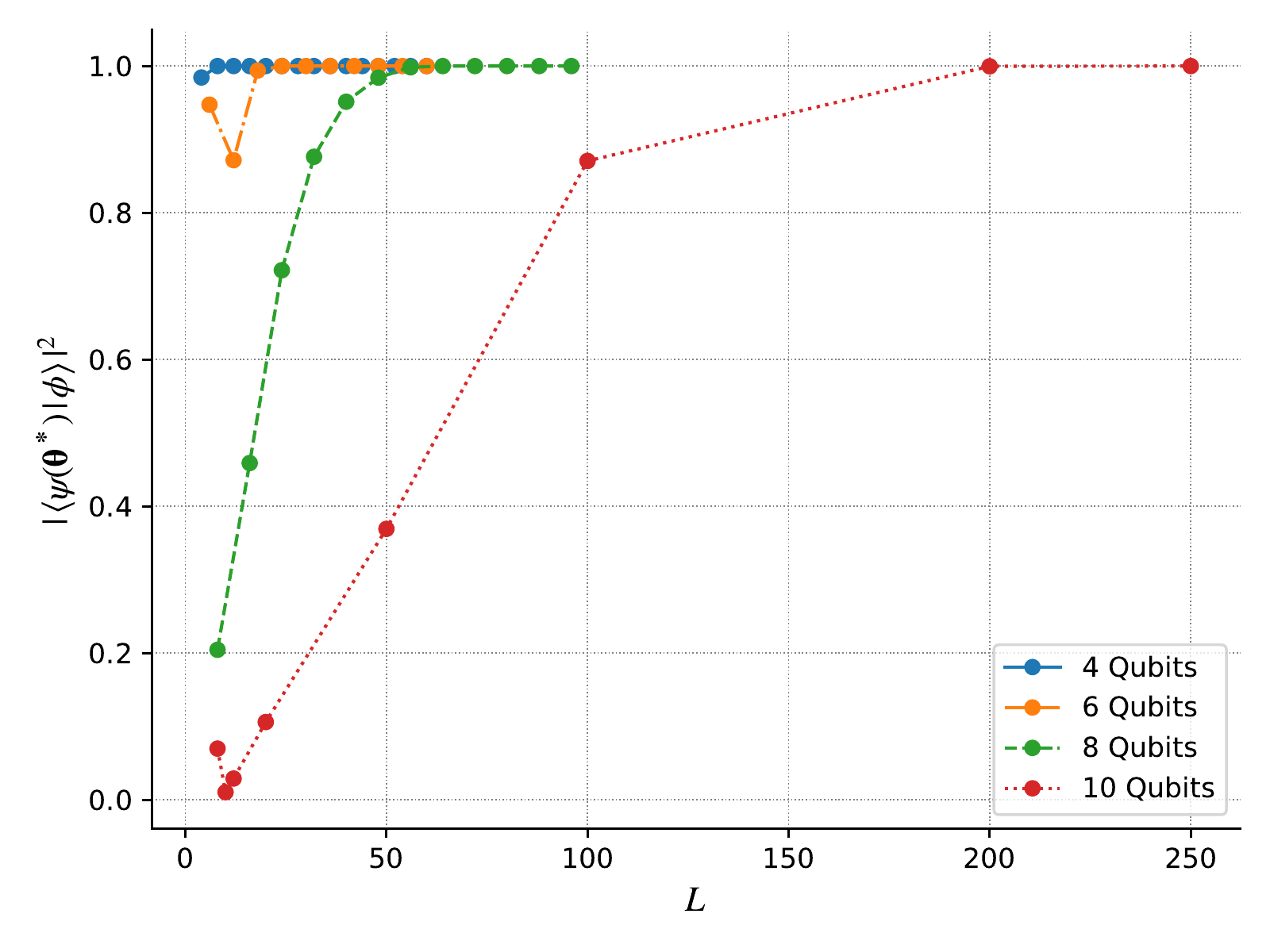}
            }
    \caption{
        Single-run VQE outcomes for the Ising model \eqref{eq:ham_ising} using the layered circuit Ansatze \eqref{eq:circuit-ansatz} at different depths $L$.
        % (a) The optimization error $\|E(\theta^*) - E_0 \|$ between the energy estimated by the circuit Ansatz at the optimized parameters $\theta^*$ and the ground state energy. 
        % (b) The fidelity $|\langle \psi({\bm \theta^*}) | \phi \rangle |^2$ between the optimized state $|\psi(\bm{\theta}^*)\rangle$ and the ground state $|\phi\rangle$.
    }
    \label{fig:ising-optimization}
\end{figure}

%%%%%%%%%%%%%%%%%%%%%%%%%%%%%%%%%%%%%%%%%%%%%%%%%%%
%% Duplicated one just for pretty layout. (BEGIN)
\begin{figure*}[t]
    \centering
    \subfloat[With the constant hyperparameters]{
        \centering
        \label{fig:ising-convergence-no-lr}
        \includegraphics[width=0.475\linewidth]{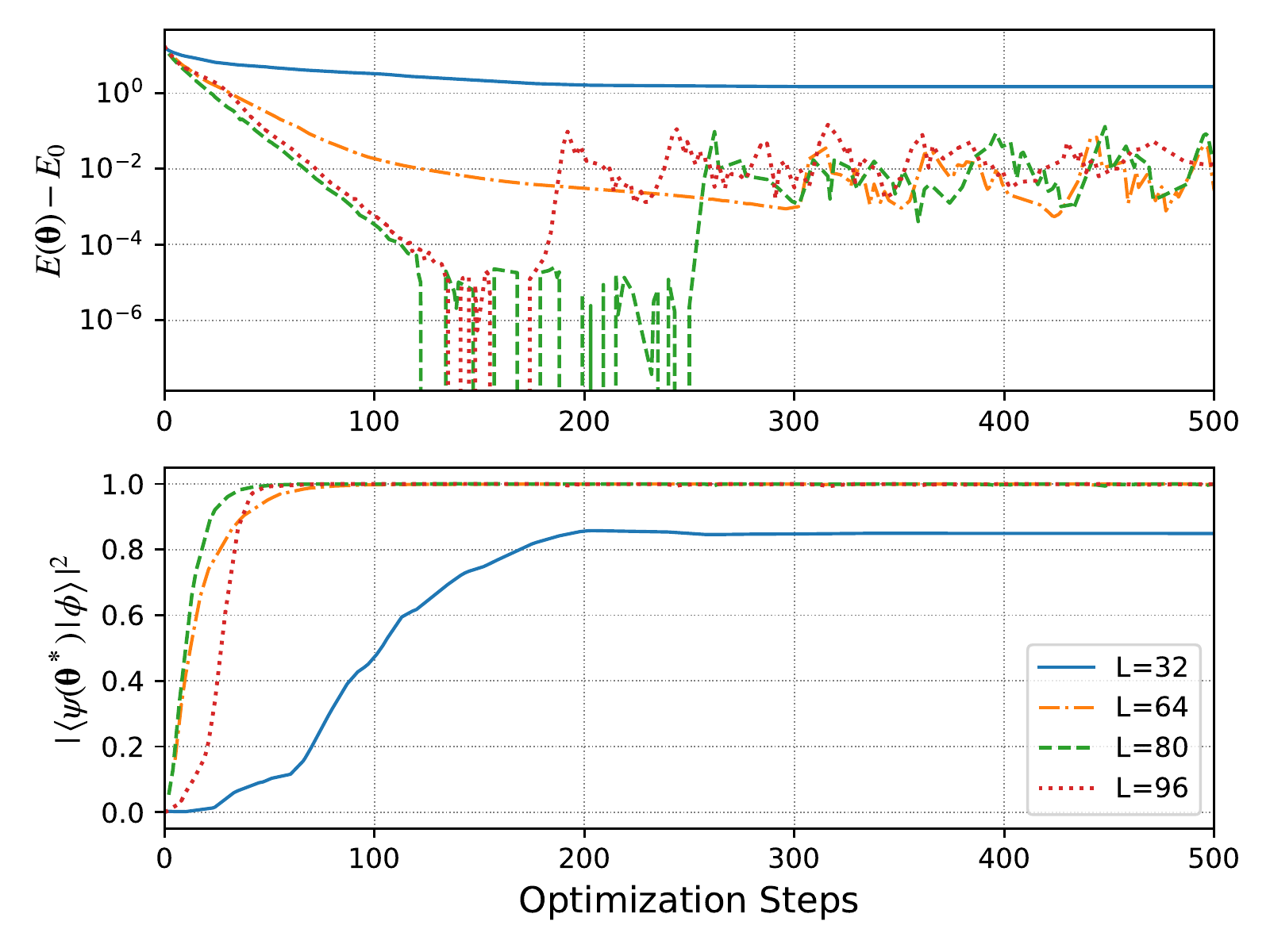}
    }
    \ \ \ 
    \subfloat[With the exponential decay \eqref{eq:lr-schedule} of hyperparameters]{
        \centering
        \label{fig:ising-convergence-lr}
        \includegraphics[width=0.475\linewidth]{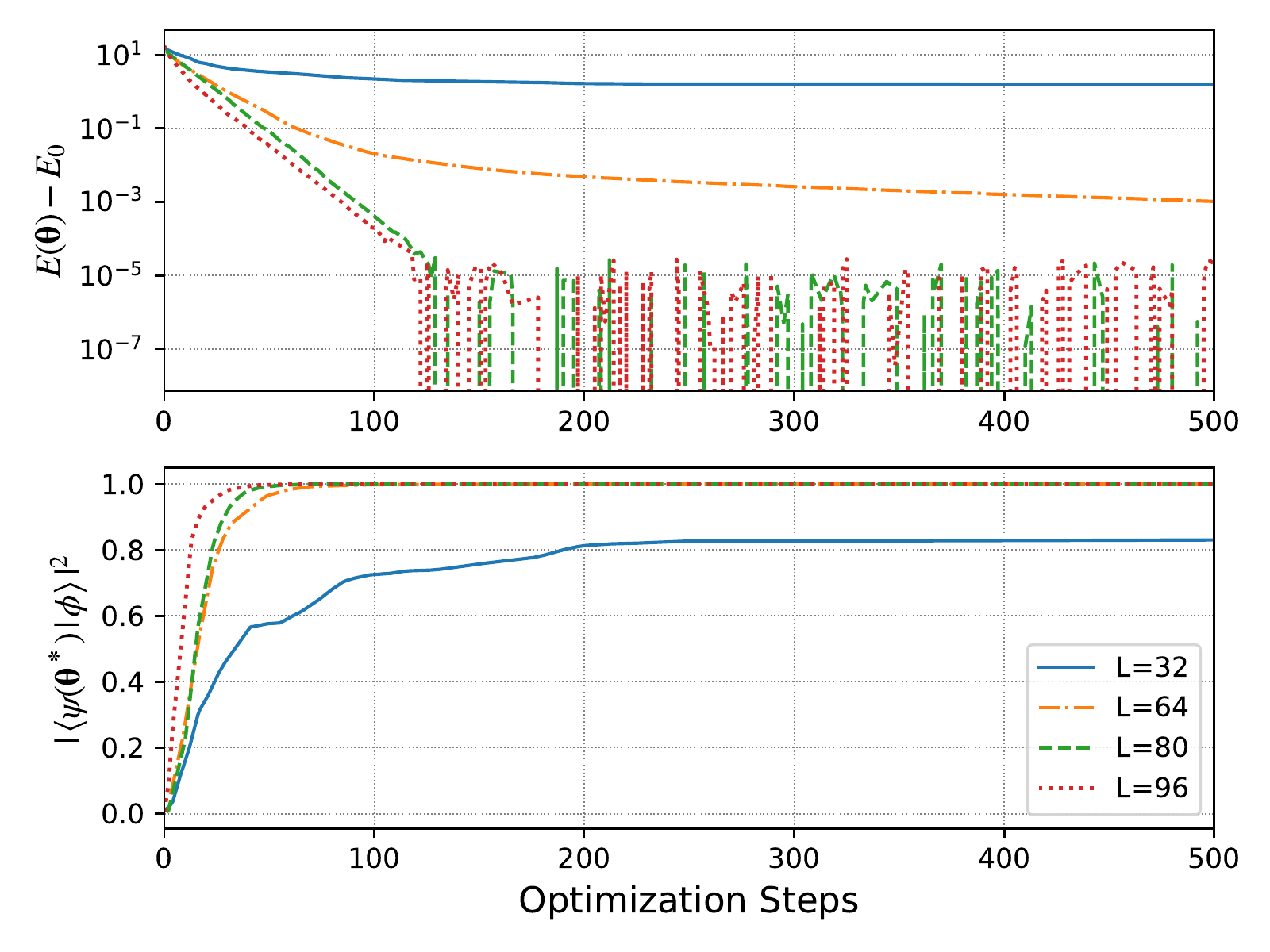}
            }
    \caption{
        Optimization curves of the circuit state on the Ising model at $8$ qubits. The upper and lower plots denote the VQE error $E(\bm{\theta}_\tau) - E_0$ and the fidelity $|\langle \psi({\bm \theta^*}) | \phi \rangle |^2$ between the circuit and true ground states, respectively, as a function of the optimization steps $\tau$. Notice the late-time fluctuation has been alleviated with the learning rate scheduling.
    }
    \label{fig:ising-convergence}
\end{figure*}

%% Duplicated one just for pretty layout. (END)
%%%%%%%%%%%%%%%%%%%%%%%%%%%%%%%%%%%%%%%%%%%%%%%%%%%

Encouraged by the observed shrinkage of the mean and variance of $E(\bm{\theta^*})$, we also have done the single VQE tryouts for a broader span of $(n, L)$, as summarized in Figure~\ref{fig:ising-optimization}. It is clear that being deeper enhances the variational circuit's capability to replicate the ground state $\vert \psi_0 \rangle$, reaching the ground state energy $E_0$ and achieving a high level of fidelity with $\vert \psi_0 \rangle$. 
%The overall characteristics of Figure~\ref{subfig:ising-error} shows is similar to Figure~\ref{fig:expressibility_mean}, 
The high-depth circuits can achieve a zero error with the following precision:
\begin{align}
\label{eq:VQE-error-bound}
    |E(\bm{\theta}) - E_0| \leq 10^{-5} \cdot \Delta E
\end{align}
where $\Delta E$ denotes the bandwidth of the target Hamiltonian $\mathcal{H}$, defined as the difference between the largest and smallest eigenvalues of $\mathcal H$. %and the ground state energy, $E_0$. 
We note that such precision can be achieved only with an appropriately chosen learning rate $\alpha$, in order to avoid too-large parameter updates that prevent the fine-level optimization. Here we  refrain from the systematic hyperparameter search, which may be more relevant for the case where the average gap between nearby energy levels shrinks, but simply stick to $\alpha = 0.05$ ($L=4, 6, 8$) and $\alpha=0.01$ ($L=10$). 
We have found that \eqref{eq:VQE-error-bound} can be achieved when the circuit depth $L$ passes through the following 
threshold value:
\begin{align*}
    \begin{array}
    {>{\raggedleft}p{1.65cm}|*{4}{>{\centering}p{0.75cm}}}
    \hline \text{$n$ qubits } & 4 & 6 & 8 & 10 \cr
    \hline \text{$L_{v}$ layers } & 10 & 24 & 68 & 250 \cr\hline
    \end{array}
\end{align*}

We also remark that deeper circuits do not necessarily lead to better performance with VQE, as displayed by the gentle ramp after passing the threshold point $L_v$. This can be understood as follows: As the space of variational parameters $\bm{\theta}$ has more dimensions, the basin of attractor to local extrema becomes narrower \cite{2020COST_BARR}, giving a larger value of the estimated inverse volume \cite{BasinAttractor2017}
\begin{align}
    \label{eq:attractor_volume}
    V_k^{-1} = \sum_{i=1}^k \log{\lambda_i(H)} 
\end{align}
where the summation is taken over the top-$k$ eigenvalues $\{{\lambda_i(H)} \}_{i=1}^k$ of the Hessian matrix $H_{ij} \equiv \partial_{i}\partial_j E({\bm{\theta})}$. For instance, the positive correlation between 
$L$ and $V_{k}^{-1}(L)$ can clearly be identified in Figure~\ref{subfig:hess-k}, drawn for  $k=100$.
As a result, the VQE trajectory is unlikely to land at an exact extremum but wander around nearby points, whose deviation gets larger as the attractor basin becomes narrower and steeper.

% \begin{figure*}[t]
%     \centering
%     \subfloat[With the constant hyperparameters]{
%         \centering
%         \label{fig:ising-convergence-no-lr}
%         \includegraphics[width=0.475\linewidth]{fig/ising_optimization.pdf}
%     }
%     \ \ \ 
%     \subfloat[With the exponential decay \eqref{eq:lr-schedule} of hyperparameters]{
%         \centering
%         \label{fig:ising-convergence-lr}
%         \includegraphics[width=0.475\linewidth]{fig/ising_optimization_ed.pdf}
%             }
%     \caption{
%         Optimization curves of the circuit state on the Ising model at $8$ qubits. The upper and lower plots denote the VQE error $E(\bm{\theta}_\tau) - E_0$ and the fidelity $|\langle \psi({\bm \theta^*}) | \phi \rangle |^2$ between the circuit and true ground states, respectively, as a function of the optimization steps $\tau$. Notice the late-time fluctuation has been alleviated with the learning rate scheduling.
%     }
%     \label{fig:ising-convergence}
% \end{figure*}

Even with the optimal number of layers, such that the circuit state can reach an exact extremum and accurately represent the ground state $\vert \psi_0\rangle$ of the Ising Hamiltonian \eqref{eq:ham_ising}, the narrowness of the attractor basin still makes the VQE trajectory somewhat unstoppable, passing through the best point $\bm{\theta^*}$ and then hopping around in the local neighborhood. Figure~\ref{fig:ising-convergence-no-lr} shows that the VQE error  $E({\bm{\theta}_\tau}) - E_0$
slightly increases on average and mildly fluctuates after achieving the minimum error $E(\bm{\theta^*}) - E_0$. This residual error can be reduced by making use of popular optimization tricks, such as early stopping or learning rate scheduling, which causes the VQE optimization to stop at the optimal point. For instance, by introducing the exponential decay of the learning rate $\alpha$, \textit{i.e.}
\begin{align}
\label{eq:lr-schedule}
    \alpha_{\tau} = \alpha_0 c^{\tau/500},\quad \tau \geq 0
\end{align}
at the optimization step $\tau$ with a constant value $c=0.3$, we could reduce the late time fluctuations as in Figure~\ref{fig:ising-convergence-lr}.

So far, we have discussed some important aspects of the VQE optimization. The main observation is that, when supported by the high-dimensional parameter space, randomly initialized variational circuits can approximate the Ising  ground state with remarkably high accuracy.
The energy gradient neither vanishes nor randomly fluctuates along the optimization trajectory, making it quickly converge to a local minimum that exhibits a small energy gap from the ground energy $E_0$ and high fidelity with the exact ground state $|\psi_0\rangle$. We will explore this efficiency of the high-depth circuit in some details by visualizing the VQE trajectory on the energy landscape.

\subsection{Visualizing the Trajectory}
\label{subsec:landscape}

%%%%%%%%%%%%%%%%%%%%%%%%%%%%%%%%%%%%%%%%%%%%%%%%%%%
%% Duplicated one just for pretty layout. (BEGIN)
\begin{figure*}[t]
    \centering
    \subfloat[
        \label{subfig:hess-k}
        The mean of the top $100$ Hessian eigenvalues
    ]{
        \centering
        \hspace{-0.7cm}
        \includegraphics[height=5.8cm]{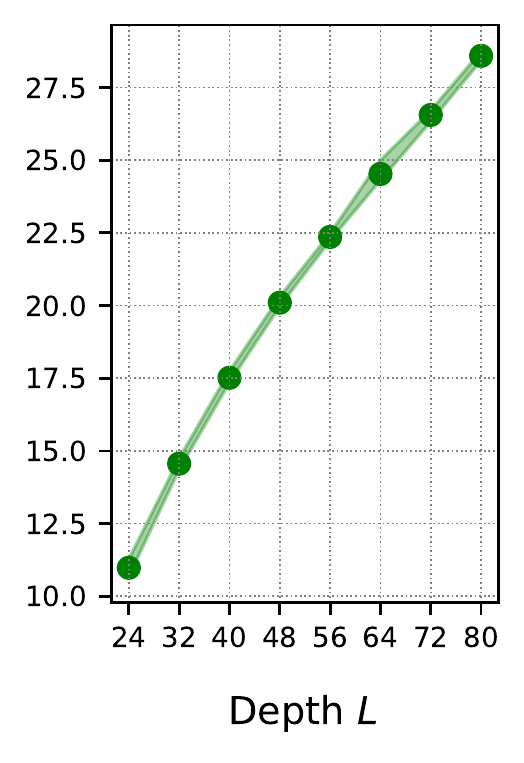}
    }    
    \subfloat[
        \label{subfig:trajectory}
        % $n=8$ and $L=24$
        The horizontal and vertical axes denote the $\mathcal{S}_{100}(\bm{\theta^*})$-projected Euclidean distance and the VQE  error $E (\bm{\theta}_\tau) - E_0$, respectively. The bottom boxes magnify the trajectory in the vicinity of the optimal point $\bm{\theta}^*$.
    ]{
        \centering
        \includegraphics[height=5.8cm]{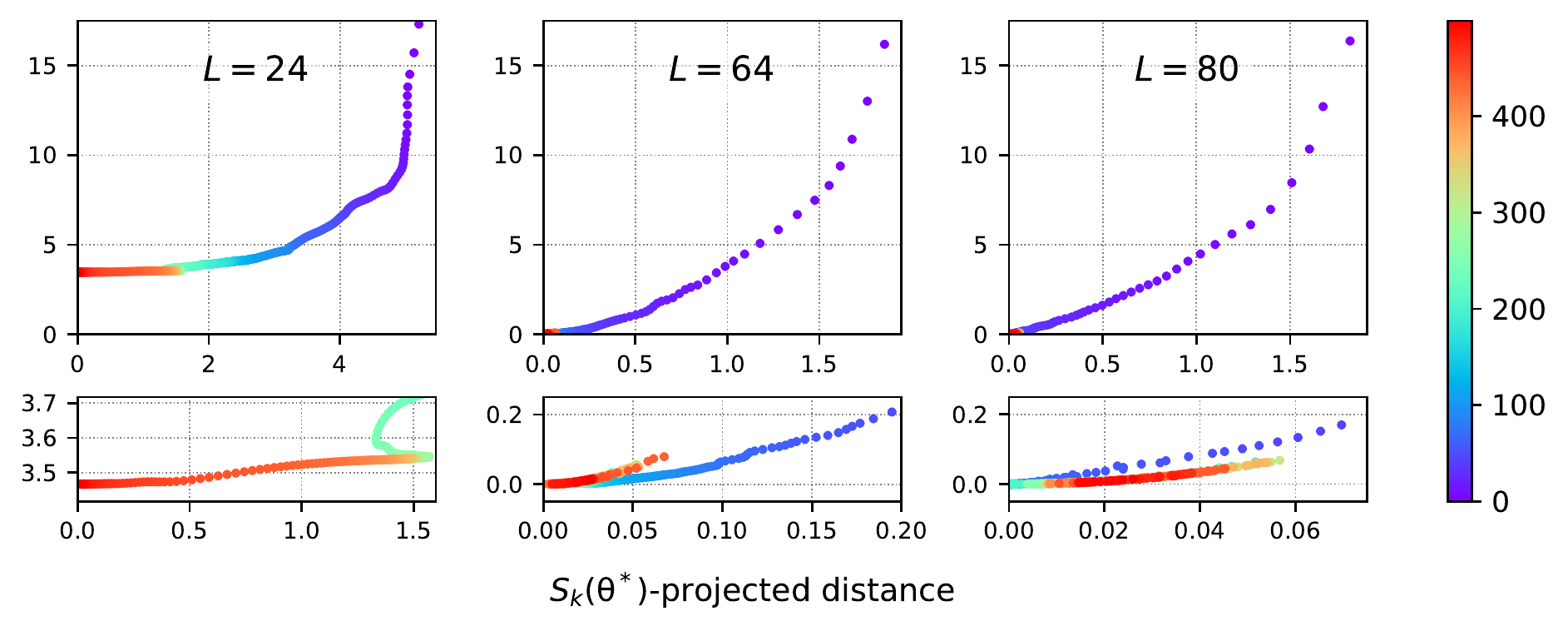}
    } 
    \label{fig:hess_eigval}
    \caption{
        The visualization of the optimization trajectory in the 100 steepest directions for the Ising model with $n=8$ qubits.
    }
    \label{fig:ising-basin}
\end{figure*}
%% Duplicated one just for pretty layout. (END)
%%%%%%%%%%%%%%%%%%%%%%%%%%%%%%%%%%%%%%%%%%%%%%%%%%%

A key characteristic that contributes to the success of the high-depth circuit is the fact that randomly initialized points are likely to be already confined in the basin of attraction of a good attractor, \textit{i.e.} a local extremum that is close enough to the ground state energy. We illustrate now this point by looking at the actual optimization trajectories under the energy minimization. 

For a given initial point $\bm{\theta}_0$ and the trajectory $\mathcal{T}(\bm{\theta}_0) = \{\bm{\theta}_\tau\}$ thereafter, we identify the optimal parameter $\bm{\theta^*}$ as the point in the trajectory $\mathcal{T}(\bm{\theta}_0)$ having the minimum energy expectation value, 
\begin{align}
    \label{eq:VQE-inequality}
    E(\bm{\theta^*})  \leq E (\bm{\theta}_\tau) \quad\text{for any }  \tau \geq 0,
\end{align}
which is the best possible representative point of the local extrema that the trajectory $\mathcal{T}(\bm{\theta}_0)$ converges around. The associated basin of attractor can be estimated by calculating the Hessian $H_{ij} \equiv \partial_{i}\partial_j E(\bm{\theta^*})$ at the optimum $\bm{\theta^*}$. 
We are interested in the eigenvalue spectrum, $\{\lambda_i(H)\}$, from which we distinguish steep and flat directions and calculate the degree of steepness. 

Similar to the case of deep neural networks \cite{Hessian2016,Valley2016, BasinAttractor2017}, the local extrema in the VQE energy landscape of high-depth circuits are often interconnected in multiple flat directions, whose corresponding Hessian eigenvalues are zero, looking like valleys rather than isolated singular points. 
Success of the VQE algorithm is not affected by a specific position in flat directions, but only concerned with if a ball, initially at a considerable height, can roll off in steep directions and reach a sufficiently deep gorge. It motivates us to consider the $k$-dimensional hypersurface $\mathcal{S}_k(\bm{\theta^*})$ along the $k$ steepest directions, \textit{i.e.}, spanned by the top-$k$ Hessian eigenvectors. More precisely, we want to examine if the $\mathcal{S}_k(\bm{\theta^*})$-projected Euclidean distance between $\bm{\theta^*}$ and $\bm{\theta}_\tau$ decreases along the trajectory $\mathcal{T}(\bm{\theta}_0)$ as the optimization step $\tau$ progresses, until 
$E (\bm{\theta}_\tau) = E(\bm{\theta^*})$. This will tell us if the entire trajectory $\mathcal{T}(\bm{\theta}_0)$ is confined in the $k$-dimensional basin of attraction, while ignoring the movement in the directions of less or zero attraction.

% \begin{figure*}[t]
%     \centering
%     \subfloat[
%         \label{subfig:hess-k}
%         The mean of the top $100$ Hessian eigenvalues
%     ]{
%         \centering
%         \hspace{-0.7cm}
%         \includegraphics[height=5.8cm]{fig/Hess_Q8_k100.pdf}
%     }    
%     \subfloat[
%         \label{subfig:trajectory}
%         % $n=8$ and $L=24$
%         The horizontal and vertical axes denote the $\mathcal{S}_{100}(\bm{\theta^*})$-projected Euclidean distance and the VQE  error $E (\bm{\theta}_\tau) - E_0$, respectively. The bottom boxes magnify the trajectory in the vicinity of the optimal point $\bm{\theta}^*$.
%     ]{
%         \centering
%         \includegraphics[height=5.8cm]{fig/Q8L246480-proj100-rainbow.pdf}
%     } 
%     \label{fig:hess_eigval}
%     \caption{
%         The visualization of the optimization trajectory in the 100 steepest directions for the Ising model with $n=8$ qubits.
%     }
%     \label{fig:ising-basin}
% \end{figure*}

Figure~\ref{subfig:trajectory} displays the VQE error $E (\bm{\theta}_\tau) - E_0$ and the projection distance $\Vert \bm{\theta}_\tau - \bm{\theta^*} \Vert_{\mathcal{S}_{100}(\bm{\theta^*})} $ along the actual optimization trajectories, whose step number $\tau$ is indicated by color. We have made the following observations: First, when an appropriate value of $k$ is selected, 
% such that the corresponding $k$-directions are sufficiently convex, 
both the distance $\Vert \bm{\theta}_\tau - \bm{\theta^*} \Vert_{\mathcal{S}_k(\bm{\theta^*})} $ and the loss value $E (\bm{\theta}_\tau) - E_0$ continuously decrease on a macroscopic scale. It exhibits that the trajectory $\mathcal{T}(\bm{\theta}_0)$ converges without escaping from a specific basin of the attractor that encloses a randomly initialized point $\bm{\theta}_0$. Second, for a shallow circuit state, the optimization trajectory often makes a slight detour in some orthogonal directions. In contrast, the steady convergence occurs typically for large $L$. It 
implies that the vicinity of any randomly initialized parameters is effectively convex.
Finally, we find from Figure~\ref{subfig:hess-k}  that the attractor basin in the direction of $\mathcal{S}_k(\bm{\theta^*})$ evolves rapidly steeper and narrower as the depth $L$ increases, thereby causing the rapid convergence and substantial late-time fluctuation around $\bm{\theta^*}$. It is noticeable that 
the initial convergence follows a milder path than later fluctuations in the $L=64$ case, while the convergence in the $L=80$ case happens along a much steeper route.
Taken together, these observations show the quick convergence of high-depth circuits \cite{2020UNITARY, 2020HVA} under the VQE algorithm.

% \begin{figure}[t]
%     \centering
%     \includegraphics[width=0.9\linewidth]{fig/ising_opt_convergence.pdf}
%     \caption{
%         (Ising)
%         The number of gradient update steps until $t^* = \min \{t\colon \|E(\theta_t) - E_0 \| < 10^{-4}\}$.
%     }
%     \label{fig:ising-convergence}
% \end{figure}

\section{Solving the SYK model}
\label{sec:syk}

Our discussions so far have been based on just a particular Hamiltonian, \eqref{eq:ham_ising}, the Ising model in a transverse uniform magnetic field.
Since the variational circuit that we use has no features particularly tailored for the Ising model, we expect it to closely replicate the ground states of other Hamiltonians in the high-depth regime. To check the generality of the prior discussions on the efficiency of the high-depth circuits, we will now solve the VQE problem, defined for another Hamiltonian of very distinct nature: the Sachdev-Ye-Kitaev (SYK) model.

\subsection{The SYK Model}
\label{subsec:SYK_intro}
The SYK model \cite{1993SY,2015Kitaev, Maldacena:2016hyu} is built out of $2n$ Majorana fermions in $1d$, \textit{i.e.} the operators $\gamma_i$, with $i = 1, \, \dots, 2n$, satisfying the following anti-commutation relations
\begin{equation}
    \left\{ \gamma_i , \gamma_j \right\} = \delta_{ij} \ ,
\end{equation}
where $\delta_{ij}$ denotes the Kronecker delta.
The SYK Hamiltonian is an all-to-all Hamiltonian, which couples all the Majorana fermions together in a fully non-local fashion, consisting of the following $q$-body interaction terms with $q\geq 2$ being an even integer:
\begin{equation}
    \label{eq:ham_syk}
    \mathcal{H} \equiv (i)^{q/2}\sum_{i_1 < \dots < i_q} J_{i_1 \dots i_q} \gamma_{i_1} \cdots \gamma_{i_q} \ ,
\end{equation}
where the coupling constants $J_{i_1 \dots i_q}$ are randomly sampled from the Gaussian distribution of  mean $0$ and variance
\begin{equation}
    \langle J_{i_1 \dots i_q}^2 \rangle \equiv \frac{J^2 (q - 1)!}{(2n)^{q - 1}} \ ,
\end{equation}
where $J^2$ is a constant which we set to be equal to one.
The model has recently attracted a widespread attention from  different communities, due to some  peculiar features it enjoys.
It has been shown that when $q \geq 4$ the model is highly chaotic \cite{2015Kitaev, Maldacena:2016hyu, Garcia-Garcia:2016mno, Cotler:2016fpe}, although solvable in the large $n$ limit \cite{2015Kitaev, Maldacena:2016hyu}, thus creating a perfect situation to study relevant questions on quantum chaos which are usually out of reach for other chaotic models.
Moreover, the SYK model has intriguing connections with the physics of black holes and quantum gravity,  promoting itself as an ideal candidate to address new questions on holography and the AdS/CFT correspondence.

We will focus our attention on the SYK model with $q = 4$. It has two notable features that can be a source of trouble for any eigensolver algorithms, regardless of being classical or quantum mechanical, whose goal is to reach the ground state. First, the energy spectrum of the SYK model is very dense, especially having a small energy gap between the ground state $E_0$ and all other excited states. Second, the SYK ground state is much less distinguishable from generic quantum states in the Hilbert space, supporting the volume law scaling of the entanglement entropy. Therefore, the VQE computation with the SYK model must be seen as a highly challenging benchmark \cite{kim2021entanglement}.

As another characteristic, the SYK model is known to have two-fold degenerate eigenstates, if and only if $n = 4k+2$ for any positive integer $k$ \cite{Garcia-Garcia:2016mno, Cotler:2016fpe}.
Denoting the two-fold degenerate ground states by $\vert \phi_1 \rangle$ and $\vert \phi_2 \rangle$, that we assume to be normalized and orthogonal to each other, the VQE target states for the SYK model are then given by all the possible linear combinations of the form:
\begin{align}
\label{eq:deg_ground_state_SYK}
    \vert \phi_0 \rangle = \frac{\alpha}{\sqrt{|\alpha|^2 + |\beta|^2}}  \vert \phi_1 \rangle + \frac{\beta}{\sqrt{|\alpha|^2 + |\beta|^2}}  \vert \phi_2 \rangle
\end{align}
with $\alpha$ and $\beta$ being complex numbers.
Therefore, the distance between the circuit state $|\psi(\bm\theta)\rangle $  and the closest state of the form $\vert \phi_0 \rangle$ can be measured by computing
\begin{align}
\label{eq:deg_ground_state_SYK-fidelity}
     \vert \langle \psi(\bm\theta) \vert \phi_1 \rangle \vert^2 + \vert \langle \psi(\bm\theta) \vert \phi_2 \rangle \vert^2 \leq 1
\end{align}
with the inequality which is saturated whenever $|\psi(\bm \theta) \rangle$ takes exactly the form \eqref{eq:deg_ground_state_SYK}.

We will numerically show that the high-depth quantum circuit can effectively learn the ground state of the SYK model. It will imply that even complicated systems, involving non-local interactions and a high level of entanglement, can be universally simulated through the variational circuits.

% An additional feature of the SYK model is that, depending on the value of $n$, the eigenstates can have a two-fold degeneracy or not, \cite{}.
% In the following, we will simply take into account this property by defining, when necessary, the fidelity $\vert \langle \psi(\bm\theta^\ast) \vert \phi \rangle \vert^2$ between the trial state $\vert \psi(\bm \theta) \rangle$ and the ground state $\vert \phi \rangle$ as
% \begin{align}
%     \vert \langle \psi(\bm\theta^\ast) \vert \phi_1 \rangle \vert^2 + \vert \langle \psi(\bm\theta^\ast) \vert \phi_2 \rangle \vert^2
% \end{align}
% where $\vert \phi_1 \rangle$ and $\vert \phi_2 \rangle$ are the two degenerate ground states.
% We will numerically show that an high-depth quantum circuit can effectively learn the ground state of the SYK model, thus showing that even complicated systems, involving non-local interactions and quantum chaos, can be universally studied through quantum variational algorithms.

\subsection{Optimizing the Circuit}
\begin{figure*}[t]
    \centering
   \subfloat[Sample variance of the energy derivative $\partial_k E(\bm{\theta})$]{
        \centering
        \label{subfig:bp-syk-a}
        \includegraphics[width=0.45\linewidth]{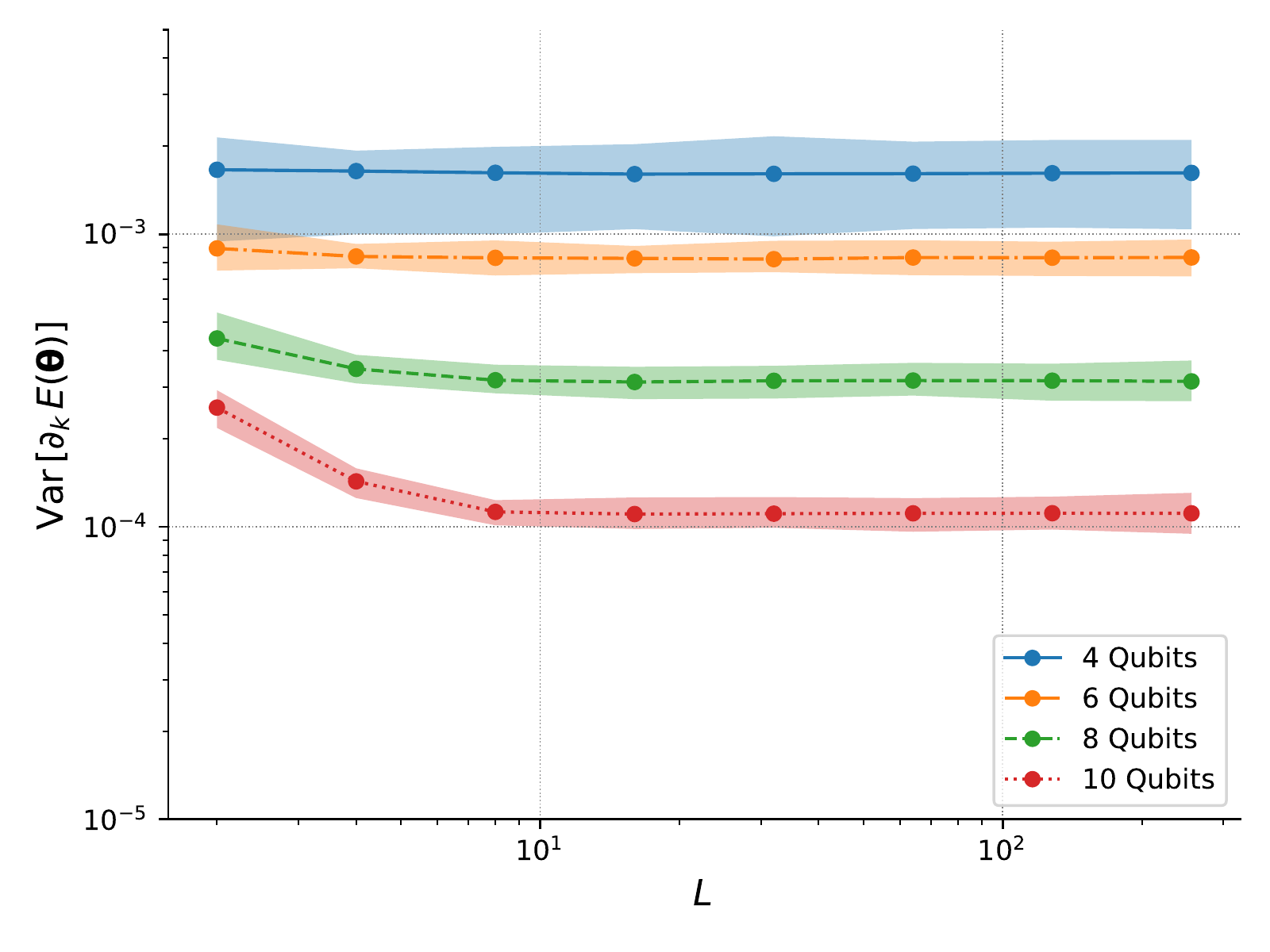}
    }
    % \newline
    \ \ \ 
    \subfloat[Sample average of the Euclidean norm of $\nabla_{\bm \theta} E(\bm{\theta})$]{
        \centering
        \label{subfig:bp-syk-b}
        \includegraphics[width=0.45\linewidth]{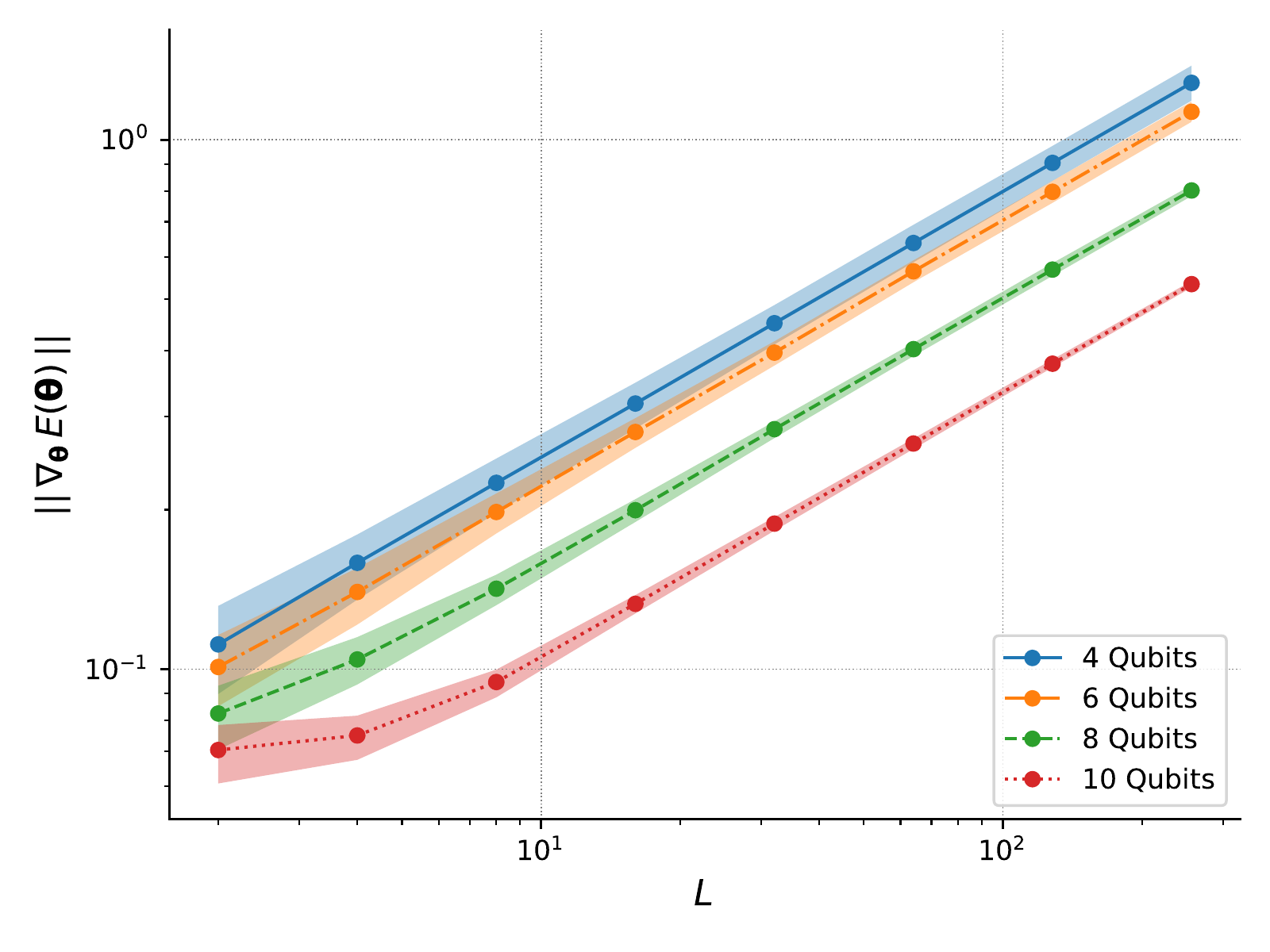}
    }
    \caption{
        The barren plateau experiment for the SYK Hamiltonian \eqref{eq:ham_syk}. The shades indicate (a) the variance across gradient components $\{\partial_k E(\bm{\theta})\}_{k=1}^{nL}$, (b) the first and third sample quantiles.
    }
    \label{fig:bp-syk}
\end{figure*}
We start by discussing if the SYK Hamiltonian causes the vanishing gradient problem for the variational circuit in Figure~\ref{fig:circuit_ansatz}.
When at least one of $U_\pm(\bm{\theta}_\pm)$  is a $2$-design, the scaling behavior of $\text{Tr}(\mathcal{H}^2)$ will determine if the SYK energy gradient will be exponentially suppressed or not.
The Hamiltonian \eqref{eq:ham_syk} has been exactly diagonalized up to $n = 15$. Moreover, an approximate analytical formula of the spectral density is at disposal in \cite{Garcia-Garcia:2017pzl}, which we use to extrapolate $\text{Tr}(\mathcal{H}^2)$ for the large system size $n$.
The results are presented in Figure~\ref{fig:spectral-density}. 
We clearly see, as in the Ising model, that  $\mathrm{Tr}(\mathcal{H}^2)$ scales like $2^n$, which indicates the abundance of the barren plateaus in the VQE energy landscape of the SYK model.

The vanishing gradient problem has also been observed numerically by computing the sample variance of $\partial_k E(\bm{\theta})$ over a collection of $1000$ random parameters, as displayed in Figure \ref{subfig:bp-syk-a}. Clearly, the energy gradient $\partial_k E(\bm{\theta})$ is exponentially suppressed by increasing $n$.
We also see that for the SYK model, contrary to the Ising Hamiltonian, there is almost no transient regime where $\partial_k E(\bm{\theta})$ is still large, yet decreasing for the growing number $L$ of layers.
%a feature that is usually seen as a motivation to use shallow circuits in the VQE algorithms \cite{2020COST_BARR}.
The lack of the transient regime, which is a consequence of the non-local nature of the SYK interactions \cite{2020COST_BARR}, is a clear obstacle in using the variational circuit to approximate the SYK ground state in the low-depth regime.

On the other hand, as we can see from Figure \ref{subfig:bp-syk-b}, the norm of the gradient vector is again increasing for the growing number $L$ of layers,
due to the saturation of the $\text{Var}_{\bm{\theta}}[\partial_{k} E(\bm{\theta})]$ with respect to $L$. The empirically measured growth rates between $\log \Vert\nabla_{\bm{\theta}} E(\bm{\theta})\Vert$ and $\log L$ are 
\begin{align*}
    \begin{tabular}
    {>{\centering}p{1.7cm}|*{4}{>{\centering}p{1.2cm}}}
        \hline
        \text{$n$ qubits} & 4 & 6 & 8 & 10 \cr
        \hline
        rate & 0.503 & 0.503 & 0.501 & 0.502 \cr
        \hline
    \end{tabular}
\end{align*}
matching the simple estimation formula \eqref{eq:rate_ising}. It tells that the gradient-based optimization can at least launch for high-depth quantum circuits,
which bypass the vanishing gradient problem using an exponentially high-dimensional parameter space.
%in the noiseless VQE setting, to replicate the ground state of the SYK model.

% and becomes of order $1$ for $L \sim 50$, thus showing that deep quantum circuits More quantitatively, we have checked the increase rate of the norm of the gradient, as we did in \eqref{eq:rate_ising}, for the SYK model
% We see that the rate is in excellent agreement with the rate computed for the Ising model.

\begin{figure}[t]
    \centering
        \includegraphics[width=\linewidth]{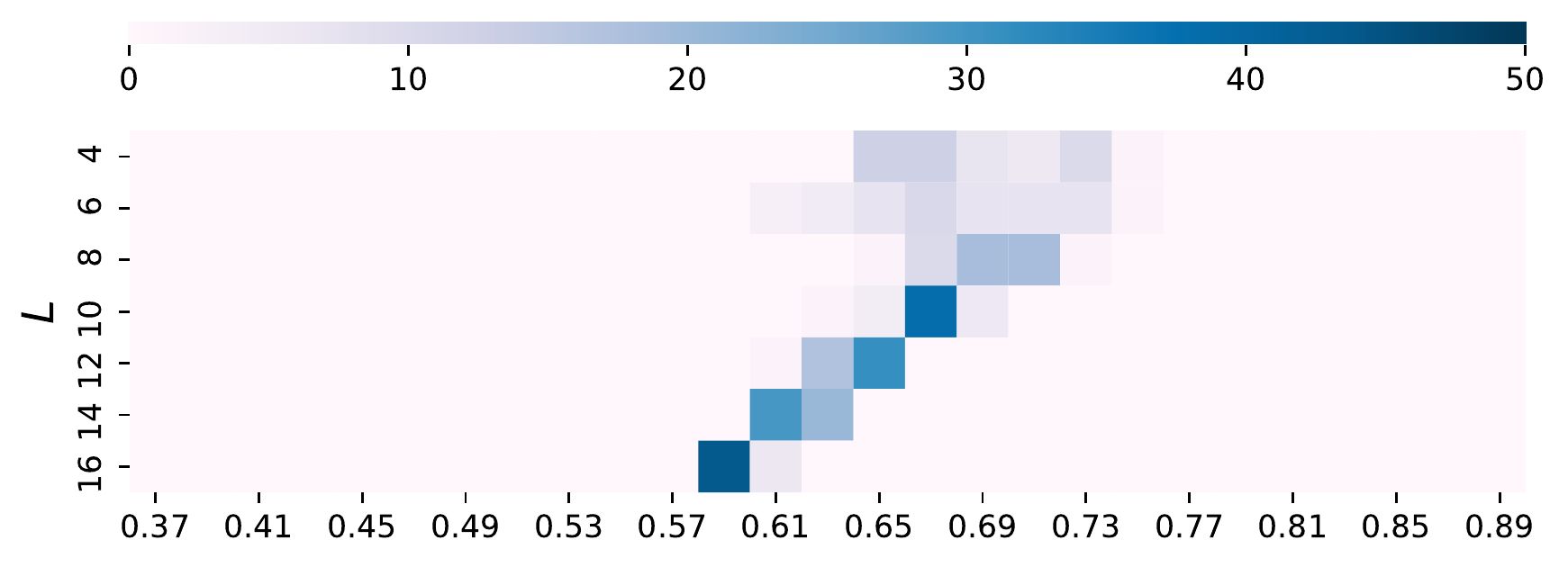}
    \caption{
        Optimized VQE energy ($E(\bm{\theta^*}) - E_0$) density for the SYK model over 35 distinct runs with random initialization.
    }
    \label{fig:vqe-syk-density}
\end{figure}

As the next step, we examine the performance of the variational circuit Ansatz \eqref{eq:circuit-ansatz} in reaching the ground state of the SYK Hamiltonian,
by repeating the same numerical experiments conducted in Section~\ref{subsec:speedup} under the same choice of hyperparameters $(\alpha, \beta_1, \beta_2)= (0.05, 0.9, 0.999)$.

First, the sample distribution of the minimized  energy $E(\bm\theta^*)$ for randomly initialized circuits with $4 \leq L \leq 16$ layers is illustrated in Figure \ref{fig:vqe-syk-density}, based on 35 sample VQE runs at each $L$. A notable  observation is the small variance of the final energy, compared to the Ising VQE energy distribution in Figure~\ref{fig:vqe-ising-density}, even at very low depth. 
% for circuits of very low-depth, of  as a function of the initial random choice of the parameters as 
We interpret it as another manifestation of the fact that the ground state that we aim to approximate is less distinguishable from generic quantum states. Hence, developing an optimization trajectory demands more classical computing power through the high-dimensional $\bm\theta$-space \cite{kim2021entanglement}.
% We interpret it as another manifestation of the fact that the ground state that we aim to approximate is less distinguishable from generic quantum states, thus developing an optimization path requires the required transition depth to an approximate 2-design is comparably low with the target Hamiltonian \eqref{eq:ham_syk}.
% that even when the circuit depth is very low, the barren plateau phenomenon for the SYK model is already clearly visible.
Another -- and perhaps the most important -- point to stress is that, the mean value of the minimized circuit energy $E(\bm\theta^*)$ decreases by stacking more layers, just like what has happened to the Ising VQE problem. It suggests that the high-depth circuits can reach a very good approximation of the SYK ground states.

\begin{figure}[t]
    \centering
    \subfloat[The VQE optimization error $E(\bm{\theta^*}) - E_0$]{
        \centering
        \label{subfig:syk-error}
        \includegraphics[width=0.9\linewidth]{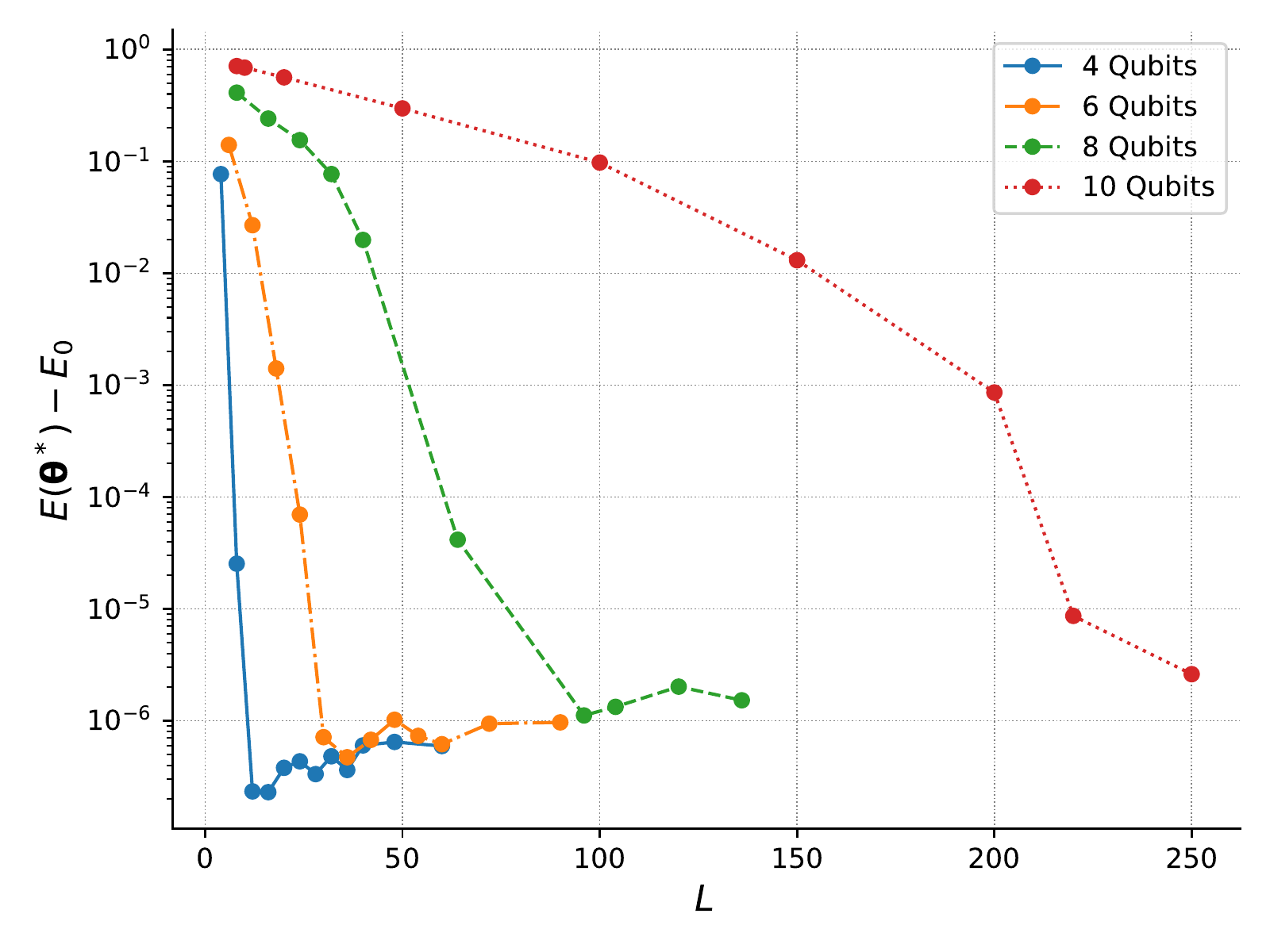}
    }
    \newline
    \subfloat[Fidelity between the optimized and true ground states. As explained near \eqref{eq:deg_ground_state_SYK}--\eqref{eq:deg_ground_state_SYK-fidelity}, for the two-fold degenerate case $n=4k+2$, the sum $|\langle \psi({\bm \theta^*}) | \phi_1 \rangle |^2 + |\langle \psi({\bm \theta^*}) | \phi_2 \rangle |^2$ is drawn.]{
        \centering
        \label{subfig:syk-fidelity}
        \includegraphics[width=0.9\linewidth]{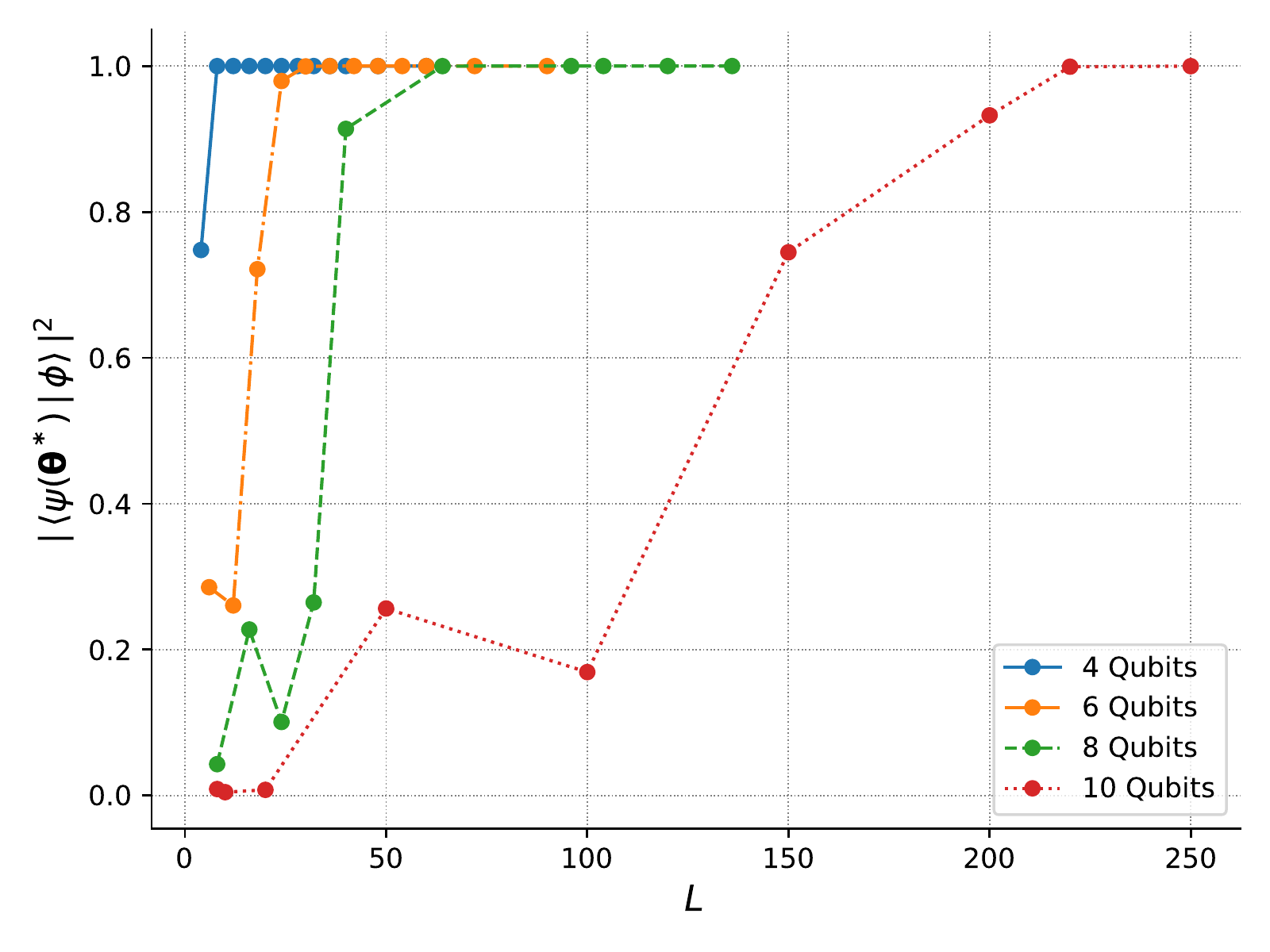}
    }
    \caption{
        Single-run VQE outcomes for the SYK model \eqref{eq:ham_syk} using the layered circuit Ansatze \eqref{eq:circuit-ansatz} at different depths $L$.
    }
    \label{fig:syk-optimization}
\end{figure}

Second, we have measured the performance of the layered circuit Ansatz \eqref{eq:circuit-ansatz} in approximating a ground state of the SYK model \eqref{eq:ham_syk}, as a function of the circuit depth.
Specifically, the VQE single-run  error $E({\bm\theta^*}) - E_0$ is drawn in Figure~\ref{fig:syk-optimization}.
The high-depth circuit performs very well, reaching a zero error with the following accuracy,
\begin{align*}
% \label{eq:VQE-error-bound}
    |E(\bm{\theta}) - E_0| \leq 10^{-5} \cdot \Delta E,
\end{align*}
when the depth $L$ arrives at the following values.
\begin{align*}
    \label{eq:vqe_syk_layers}
    \begin{array}
    {>{\raggedleft}p{1.65cm}|*{4}{>{\centering}p{0.75cm}}}
    \hline \text{$n$ qubits } & 4 & 6 & 8 & 10 \cr
    \hline \text{$L_{v}$ layers } & 12 & 30 & 96 & 220 \cr\hline
    \end{array}
\end{align*}
We also note from Figure~\ref{subfig:syk-fidelity} that the fidelity between the optimized and ground states tends to decrease at an intermediate scale of depth, while the VQE error continues to reduce without temporary increase.
Such contrasting behavior is due to the dense energy spectrum of the SYK model near the ground energy level $E_0$. With an intermediate-depth circuit, the VQE algorithm is going to approximate not the exact ground states, but some low-lying excited states. In this way, the energy continues to decrease but the fidelity does not improve.
However, for sufficiently deep circuits, the VQE algorithm can overcome the difficulty and reach an excellent agreement with the ground state. 

Interestingly, we have seen that the necessary number $L_v$ of layers to reach the high precision \eqref{eq:VQE-error-bound} is roughly in the same order, both for the SYK and Ising models. 
We will also see that the circuit \eqref{eq:circuit-ansatz} with $L \geq L_v$ achieves high precision in replicating random states by minimizing the Euclidean distance, as shown in Figure~\ref{fig:expressibility_mean}. 
This compatibility highlights the universal effectiveness of the high-depth circuit in approximating any quantum state in the Hilbert space, both  \textit{generic} and \emph{non-generic} ones.

% This number is also in rough agreement with the number of layers necessary to reach a good level of expressibility, 

% This suggests that the number of layers necessary to reach a \textit{generic} state, as measured in Figure \ref{fig:expressibility_mean}, is also of the same order of magnitude of the number of layers necessary to approximate well \textit{any} state in the Hilbert space, not necessarily generic.

\section{Approximating Random States Using Euclidean Loss Function}
\label{sec:euclidean}

Taking one step further, we will look into the high-depth circuit's capability to approximate generic quantum states $| \phi \rangle$ by minimizing the Euclidean distance.

% at the possibility of whether circuits can approximate arbitrary quantum states.
% To start with, let us evaluate the denoted by , with an appropriate choice of the circuit parameters $\bm\theta$.

% to check the expressibility of our circuit, \textit{i.e.} its ability to approximate well random states, 

% Such a map is called to be a $t$-design if the statistical distribution of the circuit states mimics the distribution of the Haar random states up to the $t$'th-order moments \cite{2007TDESIGN}. For verification of their $t$-designness, it is required to compute the $t$'th frame potential of the Ansatz state distribution, 
% \begin{align}
%     \label{eq:frame-potential}
%     \mathcal{F}^{(t)} = \int d\bm{\theta} \int d\bm{\varphi}\ |\langle \psi({\bm{\theta}})|\psi({\bm{\varphi}}) \rangle|^{2t} ,
% \end{align}
% and check if it saturates the lower bound of the inequality,
% \begin{align}
%     \mathcal{F}^{(t)} \geq  \mathcal{F}^{(t)}_{\text{Haar}} = \frac{t! (2^n - 1)!}{(t + 2^n - 1)!},
% \end{align}
% which is the $t$'th frame potential of the Haar state distribution  \cite{Welch_1974,Scott_2008,2019Expressibility}. 
% However, the Monte Carlo integration of \eqref{eq:frame-potential} can be computationally demanding due to the need for collecting exponentially many samples with respect to the number of qubits $n$ and layers $L$, also known as the curse of dimensionality.

Recall that the variational circuit $|\psi(\bm{\theta})\rangle$ with $L$ layers depends on $nL$ parameters, 
%$\bm{\theta} = \{\theta_{(i-1)n + a} \,|\, 1 \leq i \leq L$ and $1 \leq a \leq n\}$, 
which are periodic over the finite interval $\left[0, 2\pi\right)$ up to an overall sign. It is a mapping from the $nL$-dimensional torus to the Hilbert space of $n$ qubits.
% Instead, 
As a practical measure for
% more practical measure of 
the expressibility and trainability of the circuits in simulating quantum states,
% in the HQC algorithms, 
we consider whether, for a given quantum state $|\phi\rangle$, there exist a point $\bm{\theta}^*$ in the parameter space, reachable by the gradient-based optimization such that $|\psi(\bm{\theta}^*)\rangle \simeq |\phi\rangle$. Such measure of expressibility is motivated by \cite{2019Expressibility} but tailored for the hybrid algorithms. Its definition follows:

Let us apply the gradient descent to find the minimum distance at the closest point
\begin{align}
    \bm{\theta}^* = \arg\min_{\bm{\theta}} \Vert  |\psi({\bm{\theta}}) \rangle - |\phi \rangle \Vert 
\end{align}
between the circuit and target states, where $\Vert \cdot \Vert$ is the Euclidean norm of a complex vector. The (in)expressibility of the variational circuit is written as the average of the minimum distances over all quantum states:
\begin{align}
    \label{eq:expressibility}
    \varepsilon = \frac{1}{\text{Vol($U(2^n)$)}}\int d\mu_\phi\, \min_{\bm{\theta}} \Vert  |\psi({\bm{\theta}}) \rangle - |\phi \rangle \Vert.
\end{align}

Notice that the closeness between two quantum states is usually defined by using the trace distance, which is highly sensitive to small parameter changes.
Instead, we have adopted the Euclidean distance to improve the convexity of the optimization landscape, such that the closest point $\bm{\theta}^*$ can be found as easily as possible.
% , trying to unentangle the notion of the parametric (in)expressibility from actual geometric obstacles in applying HQC algorithms \cite{McClean2018BarrenPI,2020COST_BARR}. 
The Euclidean norm defines trivially a convex landscape. Hence, any non-convexity of the optimization landscape is inherited from the variational circuit itself that fixes how to embed the $nL$-dimensional torus into the Hilbert space.

% $|\psi({\bm{\theta}^*}) \rangle$ and  $|\phi \rangle$, where $\Vert |v\rangle \Vert$ is defined by the Euclidean norm of a complex vector $|v\rangle$. 

% To have a quicker check of the expressibility of our circuit ansatz, we find it sufficient to examine the parametric expressibility of a state ansatz $|\psi(\bm{\theta})\rangle$, i.e., if 

% whose integral must be practically substituted with the sample mean $\varepsilon_m$ over $m$ random states.
% We numerically assess the parametric (in)expressibility of the circuit Ansatzes \eqref{eq:circuit-ansatz} with different numbers of layers $L$, estimating the above mentioned measure \eqref{eq:expressibility} with the sample mean  over $m$ random states. 

For an actual estimation of $\varepsilon$, we 
%For the actual benchmark of the layered circuit's parametric expressibility, with different numbers $L$ of layers, it is more convenient to 
substitute the Haar unitary integral \eqref{eq:expressibility} with the sample mean over $m$ states:
\begin{align} \label{eq:expressibility_sample_mean}
    \varepsilon_m = \frac{1}{m} \sum_{i=1}^m \min_{\bm{\theta}} \Vert  |\psi({\bm{\theta}}) \rangle - |\phi_i \rangle \Vert 
\end{align}
Given a target state $\vert \phi_i\rangle$, we start with an Ansatz state $|\psi({\bm{\theta}}) \rangle $ with $nL$ initial parameters $\bm{\theta}$, randomly sampled from the uniform distribution $\mathcal{U}(0, 2\pi)^{\otimes nL}$.
To reach the optimal parameters $\bm{\theta}^* $ that minimize the distance, we use the Adam optimization algorithm with the hyperparameters  $(\alpha, \beta_1, \beta_2)= (0.05, 0.9, 0.999)$. 
Note that the overwhelming majority of Hilbert space is filled with generic states with a high degree of entanglement, so the sample states $\vert \phi_i\rangle$ will almost never be non-generic, low-entangling states similar to the Ising ground state.

% \cite{2015Adam}, 
% which iteratively updates the variational parameters $\bm{\theta}$ by the exponential moving averages of gradients and their squares. It has a clear advantage in convergence speed, being widely used in a variety of deep learning models such as recurrent networks \cite{2017Attention,2019BERT}, convolution networks \cite{2017SuperResolution}, and graph networks \cite{2017GCN}. The parameter update rule is concretely determined by a choice of the hyperparameters $(\alpha, \beta_1, \beta_2)$ defined in \cite{2015Adam}. 
% In the current benchmark, we have specifically used the learning rate $\alpha = 0.05$ and the exponential decay rates $\beta_1=0.9$ and $\beta_2=0.999$ for the moving averages. 

\begin{figure}[t]
    \centering
    \includegraphics[width=\linewidth]{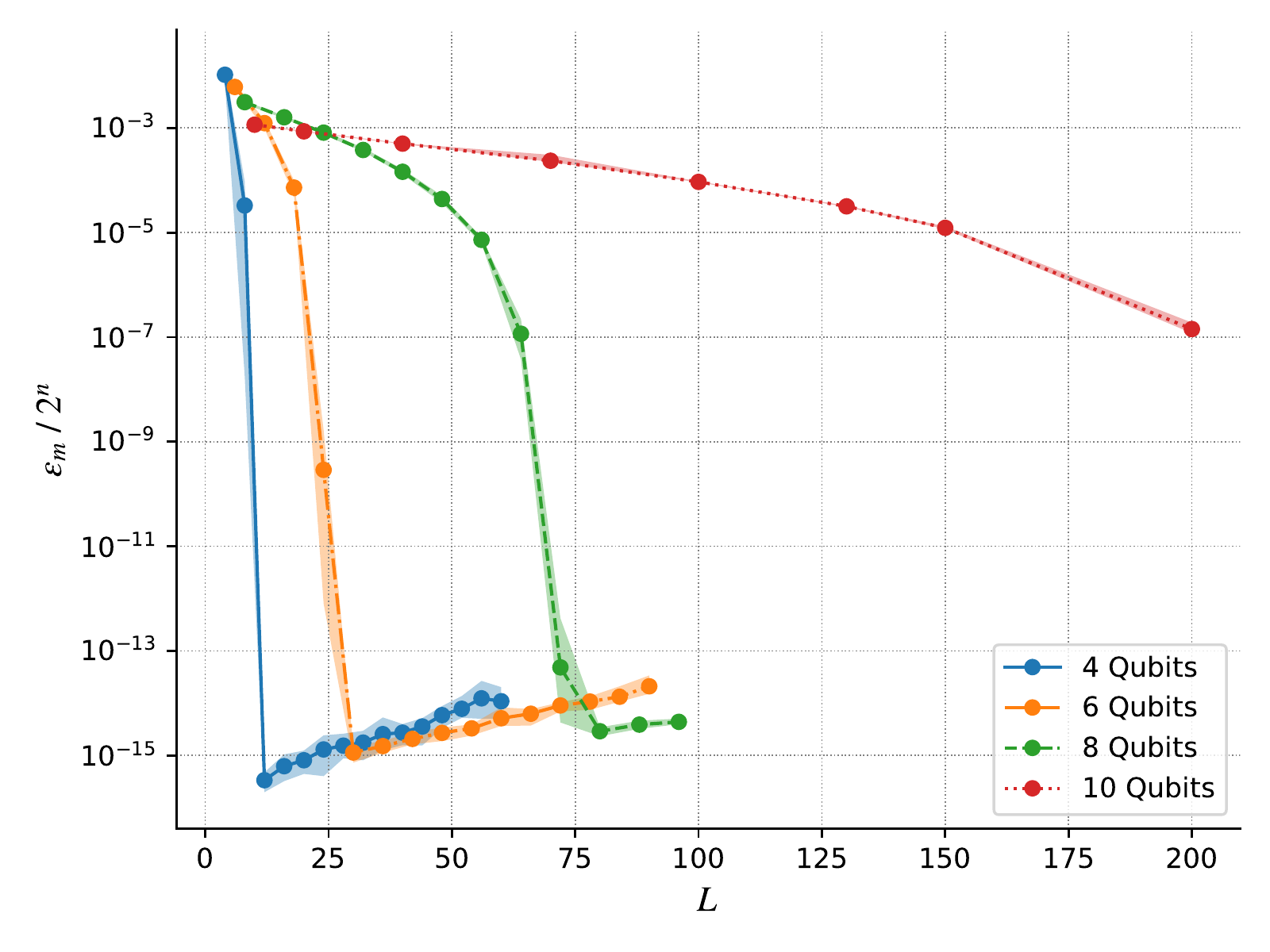}
    \caption{
        The parametric (in)expressibility of the layered circuit Ansatz over $m=10$ random target samples, $\varepsilon_{m=10} / 2^n$, normalized by the number $2^n$ of state components. The narrow shade denotes the fluctuation across $m=10$ target states.
    }
    \label{fig:expressibility_mean}
\end{figure}

% For each case we took the parameter $\hat{\theta}$ that minimizes the cost function (\ref{eq:expressibility_cost}) among the first  .

Figure \ref{fig:expressibility_mean} shows the sample mean $\varepsilon_m$ over $m=10$ random target states, divided by the number $2^n$ of state components. The narrow shade displays the fluctuation range of the component-averaged distance $\| | \psi(\bm{\theta^*}) \rangle - | \phi_i \rangle \|  \,\cdot\, 2^{-n}$ across different target states indexed by $i=1,\ldots,10$. We have selected the minimum distance among the first $\tau\leq 500$ optimization steps, which is sufficient since the optimal parameter $\bm{\theta}^*$ is empirically always found in an early stage of the optimization.

We find that the (in)expressibility approaches to 0 as the circuit depth $L$ grows. To achieve the following high precision, 
\begin{align}
    \label{eq:expressible_bound}
    \varepsilon_{m=10} \leq 10^{-5} \cdot 2^n,
\end{align} 
the depth $L$ has to be greater than the threshold values:
\begin{align*}
    % \label{eq:expressible_layers}
    \begin{array}
    {>{\raggedleft}p{1.65cm}|*{4}{>{\centering}p{0.75cm}}}
    \hline \text{$n$ qubits } & 4 & 6 & 8 & 10 \cr
    \hline \text{$L_{\varepsilon}$ layers } & 10 & 24 & 56 & 150 \cr\hline
    \end{array}
\end{align*}
Note that the slight ramp-up after the steep fall of $\varepsilon_{m=10}$ is caused by the fluctuation of $\bm{\theta}$ around the optimal point $\bm{\theta^*}$, since the fluctuation size gets amplified for bigger $L$. It is a common phenomenon in gradient-based optimization, where a suitable reduction of the learning rate can turn it to the stable convergence $\bm{\theta} \rightarrow \bm{\theta^*}$.
% For instance, at the learning rate $\alpha=0.01$, the 6-qubit circuit $| \psi(\bm{\theta}) \rangle$ with $L=90$ layers achieves $\varepsilon_{m=10} < 10^{-15} \cdot 2^6$, which is roughly the same as of $L=24$ layers.

% By evaluating the layered circuit Ansatz \eqref{eq:circuit-ansatz} of variable depths, based on their parametric (in)expressibility $\varepsilon$, 
In Figure~\ref{fig:expressibility_mean},
we have found the following overall trend: Deeper circuits are not only a superset of shallow circuits but also behave more effectively in the gradient-based optimization. Though the functional form of the variational circuit in Figure~\ref{fig:circuit_ansatz} is composed of only two types of gates, it can accurately express and reach most quantum states in Hilbert space in the high-depth regime.

\section{Discussion}
\label{sec:discussion}

From the viewpoint of the vanishing gradient problem, adding more parametric gates to the circuit architecture brings both positive and negative effects on trainability. On the negative side, it makes the circuit unitary ensemble closer to quantum 2-design, whose one- and two-point correlators are equivalent to those of Haar random unitaries, so that initial random gradients may decay exponentially with the system size $n$ \cite{McClean2018BarrenPI}. On the positive side, however, it enlarges the dimensionality of the parameter space and maintains the norm of random gradients to be finite. 

Until now, we have explored one limiting regime, where the impact of exponentially many parameters dominates over the barren plateau phenomenon. 
It turns out that the high-depth circuit has been very effective in solving the ground states of the Ising and SYK Hamiltonians, as well as in simulating generic random states.
One may note a certain degree of qualitative similarity between the VQE optimization trajectory of high-depth circuits, demonstrated in Section~\ref{subsec:speedup}, and the lazy learning \cite{2018LAZY_LEARNING} in over-parameterized neural networks. We speculate it as naturally emerging in any systems involving the high-dimensional parameter space. It would be  interesting to know why local extrema on the energy landscape of the high-depth circuits can reliably reach a zero VQE error, as studied in \cite{2016NO_BAD_LOCAL_MINIMA} for the neural networks.

Some interesting phenomena that we found during the optimization of low- and intermediate-depth circuits need to be further understood.  For low-depth circuits, it is important to characterize what features of the initial points lead to the observed big difference in their minimized energy levels. We are also curious to know what makes the local minima on the energy landscape of intermediate-depth circuits so homogeneous. More generally, we want to understand how the circuit states evolve along a gradient optimization trajectory on average, in terms of various quantum information measures.

We need to concern two types of errors for the actual use of the variational circuits on the near-term quantum devices \cite{Preskill_2018}. Firstly, for accurate estimation of the energy gradient, it is necessary to sample the variational wavefunction repeatedly, ideally exponentially many times in the system size $n$. Under the assumption that we can collect only a limited number of samples, the energy gradient estimation will be inevitably noisy. Therefore, the expressibility and trainability of the variational circuit needs to be addressed with noisy gradients, or even without direct gradient computation as in  \cite{maheswaranathan2019guided,welleck2020mleguided}. We expect the simplicity of the energy landscape in the high-depth regime may bring robustness against the sampling noise. 
Secondly, and more severely, quantum hardware noise restricts our ability to maintain quantum states when a sequence of layer unitary acts on them. A noise-induced mechanism that causes the vanishing gradient problem has also been studied in \cite{2020NOISE_BARR}. We would like to investigate if the variational circuit approach can be successful under decoherence of quantum states in the near future.

\begin{acknowledgments}
We thank Boris Hanin and Yaron Oz for helpful discussions. 
Joonho Kim acknowledges the support from the NSF grant PHY-1911298 and the Sivian fund. The work of DR is supported by the National Research Foundation of Korea (NRF) grant NRF-2020R1C1C1007591. 
Our Python scripts, which are available \href{https://github.com/phyjoon/high-depth-vqe}{https://github.com/phyjoon/high-depth-vqe}

for the numerical experiments written in \texttt{JAX} \cite{2018JAX},  \texttt{QuTip} \cite{2013QUTIP}.
% and can be found on \url{https://github.com/.../}. 
The experimental data are managed in \href{https://www.wandb.com}{Weights~\&~Biases} \cite{WANDB}.

\end{acknowledgments}

\bibliographystyle{apsrev4-1}
\bibliography{reference.bib}% Produces the bibliography via BibTeX.

\end{document}